\RequirePackage{fix-cm}
\documentclass{svjour3} 
\usepackage{paralist}
\usepackage{graphics}
\usepackage{multicol}
\usepackage{multirow}
\usepackage{siunitx}
\usepackage{threeparttable}
\usepackage{pifont}
\usepackage{array}
\usepackage[table]{xcolor}
\usepackage[linesnumbered,ruled,vlined]{algorithm2e}
\usepackage{float}
\usepackage{tikz}
\usepackage{comment}
\usetikzlibrary{automata,positioning,fit,arrows.meta,backgrounds,shapes.geometric}
\usepackage{mathtools,halloweenmath}
\usepackage{booktabs}
\usepackage[numbers]{natbib}
\usepackage{hyperref}
\usepackage{subcaption}
\usepackage{makecell}
\usepackage{tcolorbox}

\journalname{Empirical Software Engineering}

\begin{document}
\title{Systematic Evaluation of Deep Learning Models for Log-based Failure Prediction\footnote{The version of record of this article, first published in Empirical Software Engineering, is available online at Publisher’s website: \url{https://dx.doi.org/10.1007/s10664-024-10501-4}}}

\author{
	Fatemeh Hadadi
        \and Joshua H.\ Dawes
        \and Donghwan Shin
        \and Domenico Bianculli
        \and Lionel Briand
}


\institute{
    Fatemeh Hadadi \at
	University of Ottawa, Canada \\
\email{fhada072@uottawa.ca}
\and
Lionel Briand \at University of Ottawa, Canada, and Lero centre, University of Limerick, Ireland (Part of this work was done while the author was with the University of Luxembourg)\\
\email{lbriand@uottawa.ca}
\and
		Joshua H.\ Dawes \and Domenico Bianculli \at
	SnT Centre, University of Luxembourg, Luxembourg \\
	\email{joshua.dawe@uni.lu, domenico.bianculli@uni.lu}
\and
	Donghwan Shin \at University of Sheffield, United Kingdom (Part of this work was done while the author was with the University of Luxembourg) \\
	\email{d.shin@sheffield.ac.uk}
}

\date{Accepted: 17 May 2024 / Published: 20 June 2024}
\maketitle

\begin{abstract}
With the increasing complexity and scope of software systems, their dependability is crucial. The analysis of log data recorded during system execution can enable engineers to automatically predict failures at run time. Several Machine Learning (ML) techniques, including traditional ML and Deep Learning (DL), have been proposed to automate such tasks. However, current empirical studies are limited in terms of covering all main DL types---Recurrent Neural Network (RNN), Convolutional Neural Network (CNN), and transformer---as well as examining them on a wide range of diverse datasets.

In this paper, we aim to address these issues by systematically investigating the combination of log data embedding strategies and DL types for failure prediction. To that end, we propose a modular architecture to accommodate various configurations of embedding strategies and DL-based encoders. To further investigate how dataset characteristics such as dataset size and failure percentage affect model accuracy, we synthesised \num{360} datasets, with varying characteristics, for three distinct system behavioural models, based on a systematic and automated generation approach. Using the F1 score metric, our results show that the best overall performing configuration is a CNN-based encoder with Logkey2vec. Additionally, we provide specific dataset conditions, namely a dataset size $>350$ or a failure percentage $>7.5\%$, under which this configuration demonstrates high accuracy for failure prediction.


\end{abstract}

\section{Introduction}\label{sec:introduction}

As software systems continue to increase in complexity and scope, reliability and availability play a critical role in quality assurance and software maintenance~\cite{bauer2012reliability, 10.1145/3510003.3510155}. During runtime, software systems often record log data about their execution, designed to help engineers monitor the system’s behaviour~\cite{He2021}. One important quality assurance activity is to predict failures at run time based on log analysis, as early as possible before they occur, to enable corrective actions and minimise the risk of system disruptions~\cite{CARVALHO2019106024}.

However, software systems typically generate a vast quantity of log data which makes manual analysis error-prone and extremely time-consuming. Therefore, a number of automatic log analysis methods, particularly for failure prediction~\cite{aarohi, Das2018, Sahoo2003} and anomaly detection~\cite{DeepLog, LogAnomaly, LogRobust}, have been proposed over the past few years. 
Machine Learning (ML) has played a key role in automatic log analysis, from Traditional ML methods (e.g., Random Forest (RF)~\cite{breiman2001random}, Support Vector Machine (SVM)~\cite{cortes1995support}, Gradient Boosting (GB)~\cite{Chen2019}) to Deep Learning (DL) methods (e.g., DeepLog~\cite{DeepLog}, LogRobust~\cite{LogRobust}, LogBERT~\cite{logBERT}) relying on various DL network architectures, including Long Short-Term Memory (LSTM), Convolutional Neural Network (CNN), and transformers~\cite{10.1145/3510003.3510155}.

Although several studies have explored the use of DL models with various log sequence embedding strategies~\cite{He2021}, they have been limited in terms of evaluating the three main types of DL networks---RNN, CNN, and transformer---combined with different embedding strategies; for instance, two studies by \citet{10.1145/3510003.3510155} and \citet{CNN} included CNN-based models but did not cover transformer-based models.
Moreover, previously studied models were often applied to a limited number of available datasets, which severely limited the generalizability of results~\cite{He2021}. Indeed, because these few datasets exhibit a limited variety of characteristics, studying the robustness and generalizability of DL models, along with their embedding strategies, is unlikely to yield practical guidelines.

In this paper, we aim to systematically investigate the combination of the main DL architectures and embedding strategies, based on datasets whose main characteristics (e.g., dataset size and failure percentage) are controlled. 
To achieve this, we first introduce a modular architecture for failure prediction, where alternative log embedding strategies and DL models can be easily applied. 
The architecture consists of two major steps: an embedding step that converts input logs into log embedding vectors followed by a classification step that predicts failures by processing the embedding vectors using encoders that are configured by different DL models, called DL encoders. 
In the embedding step, three alternative strategies, i.e., a semantic-based strategy (BERT~\cite{BERTbasemodel}), a template ID-based strategy Logkey2vec~\cite{CNN}, and aggregation of semantic and template ID-based strategies, FastText with TF-IDF~\cite{LogRobust}, are considered.
In the classification step, four types of DL models, including LSTM~\cite{LSTM}, BiLSTM\cite{BiLSTM}, CNN\cite{CNNfirstpaper}, and transformer~\cite{AttentionIA}.

Furthermore, we compared the results of our systematic investigation of DL architectures with a top traditional ML-based failure predictor to assess the advantage of DL-based approaches.

Also, to address the issue of the limited availability of adequate datasets, we designed a rigorous approach for generating synthesised data relying on behavioural models built by applying model inference algorithms~\cite{PRINS, MINT} to available system logs. 
When synthesizing data, we control key dataset characteristics such as the size of the dataset and the percentage of failures. Additionally, we define patterns that are associated with system failures and are used to classify logs for the failure prediction task. The goal is to associate failures with complex patterns that are challenging for failure prediction models. 
Further, based on our study, we investigated how the dataset characteristics determine the accuracy of model predictions and then derive practical guidelines.

Finally, we processed a real-world dataset for failure prediction, called OpenStack\_PF, to compare the results obtained on synthesized data with those obtained on a real-world failure prediction dataset. The objective was to obtain further evidence of the validity of our data synthesis strategy.

Our empirical results conclude that the best model includes the CNN-based encoder with Logkey2vec as an embedding strategy. Using a wide variety of datasets, both synthesised and real-world, showed that this combination is also very accurate when certain conditions are met in terms of dataset size and failure percentage. Our findings provide valuable insights for software and AIOps engineers to select the best DL-based solution for optimal failure prediction. 
Moreover, we aim to provide guidance in  optimising dataset characteristics to improve failure prediction accuracy. In conclusion, this paper offers clear guidelines for those looking to leverage DL in predicting system failures from logs.

To summarise, the main contributions of this paper are:
\begin{itemize}[-]
    \item A large-scale, systematic investigation of the application of various DL encoders---LSTM-, BiLSTM-, CNN-, and transformer-based---and embedding strategies---BERT~\cite{BERTbasemodel}, Logkey2vec~\cite{CNN} and hybrid strategy combining FastText with TF-IDF~\cite{LogRobust}---for failure prediction modeling
    \item A systematic and automated approach to synthesise log data, with a focus on experimentation in the area of failure prediction, to enable the control of key data set characteristics while avoiding any other form of bias.  
      \item A comparison of 
     the results obtained on synthesized data with those of a real-world dataset to provide further evidence of the validity of our data synthesis strategy.
    \item A comparison of DL-based and a best-performing traditional ML-based failure predictor to assess the benefits of the former.
    \item Practical guidelines for using DL-based failure prediction models according to dataset characteristics such as dataset size and failure rates. 
    \item A publicly available replication package, containing the implementation, generated datasets with behavioural models, and results.
\end{itemize}

The rest of the paper is organised as follows. Section~\ref{sec:background} presents the basic definitions and concepts that will be used throughout the paper. Section~\ref{sec:related-works} illustrates related work. Section~\ref{sec:architecture} describes the architecture of our failure predictor with its different configuration options. Section~\ref{sec:empirical-design} describes our research questions, empirical methodology, and synthetic log data generation. Section~\ref{sec:expr-results} reports empirical results. Section~\ref{sec:discussion} discusses the implications of the results. Section~\ref{sec:conclusion} concludes the paper and suggests future directions for research and improvements.

\section{Background}\label{sec:background}

In this section, we provide background information on the main concepts and techniques that will be used throughout the paper. 
First, we briefly introduce the concepts related to finite state automata (FSA) and regular expressions in \S~\ref{sec:background:automata} and execution logs in \S~\ref{sec:background:logs}. 
We then describe two important log analysis tasks (anomaly detection and failure prediction) in \S~\ref{sec:failurevsanomaly} and further review machine-learning (ML)-based approaches for performing such tasks in \S~\ref{sec:background:ml-based-log-analysis}. 
We conclude by providing an overview of embedding strategies for log-based analyses in \S~\ref{sec:background:embed}.

\subsection{Finite State Automata and Regular Expressions}\label{sec:background:automata}

A \emph{deterministic FSA} is a tuple $\mathcal{M} = \langle Q, A, q_0, \Sigma, \delta \rangle$, where $Q$ is a finite set of states, $A \subseteq Q$ is the set of accepting states,  $q_0 \in Q$ is the starting state, $\Sigma$ is the alphabet of the automaton, and $\delta \colon Q \times \Sigma \rightarrow Q$ is the transition function. The extended transition function $\delta^* : Q \times \Sigma^* \rightarrow Q$, where $\Sigma^*$ is the set of strings over $\Sigma$, is defined as follows:
\begin{enumerate}[(1)]
    \item For every $q \in Q, \delta^*(q,\epsilon) = q$, where $\epsilon$ represents the empty string;
    \item For every $q \in Q$, every $y \in \Sigma^*$, and every $\sigma \in \Sigma$, $\delta^*(q, y\sigma) = \delta(\delta^*(q, y), \sigma)$.
\end{enumerate}
Let $x \in \Sigma^*$; the string $x$ is accepted by $\mathcal{M}$ if $\delta^*(q_0, x) \in A$ and is rejected by $\mathcal{M}$, otherwise.

The language accepted by an FSA $\mathcal{M}$ is denoted by $\mathcal{L}(\mathcal{M})$ and is defined as the set of strings that are accepted by $\mathcal{M}$; more formally, $\mathcal{L}(\mathcal{M}) = \{w \mid \delta^*(q_0, w) \in A \}$.
A language accepted by an FSA is called a \emph{regular} language.

Regular languages can also be defined using \emph{regular expressions}; given a regular expression $r$ we denote by $\mathcal{L}(r)$ the language it represents. 
A regular expression $r$ over an alphabet $\Sigma$ is a string  containing symbols from $\Sigma$ and special meta-symbols like ``$|$'' (union or alternation), ``.'' (concatenation), and ``*'' (Kleene closure or star), defined recursively using the following rules:

\begin{enumerate}[(1)]
    \item $\emptyset$ is a regular expression denoting the empty language $\mathcal{L}(\emptyset)=\emptyset$;
    \item For every $a \in \Sigma$, $a$ is a regular expression corresponding to the language $\mathcal{L}(a)=\{a\}$;
    \item If $s$ and $t$ are regular expressions, then $r= s | t$ and $r=s.t$ (or $r=st$) are regular expressions denoting, respectively, the union and the concatenation of $\mathcal{L}(s)$ and $\mathcal{L}(t)$;
    \item If $s$ is a regular expression, then $r=s^*$ is a regular expression denoting the Kleene closure of $\mathcal{L}(s)$.
\end{enumerate}

\subsection{Logs}\label{sec:background:logs}

In general, a \textit{log} is a sequence of log messages generated by logging statements (e.g., \texttt{printf()}, \texttt{logger.info()}) in the source code~\cite{He2021}. 
A \textit{log message} is textual data composed of a \emph{header} and \emph{content}~\cite{He2021}. In practice, the logging framework determines the \emph{header} (e.g., \texttt{INFO}) while the \emph{content} is designed by developers and is composed of static and dynamic parts. 
The static parts are the fixed text written by the developers in the logging statement (e.g., to describe a system event), while the dynamic parts are determined by expressions (involving program variables) evaluated at runtime. 
For instance, let us consider the execution of the log printing statement \texttt{logger.info(``Received block\textvisiblespace"+ block\_ID)}; during the execution, assuming variable \texttt{block\_ID} is equal to \texttt{2}, the log message \texttt{Received block 2} is printed. In this case, \texttt{Received block\textvisiblespace} is the static part while \texttt{2} is the dynamic part, which changes depending on the value of \texttt{block\_ID} at run time.

A \textit{log template} (also called \textit{event template} or \textit{log key}) is an abstraction of the log message content, in which dynamic parts are masked with a special symbol (e.g., \texttt{*}); for example, the log template corresponding to the above log message is \texttt{Received block\textvisiblespace *}. Often, each unique log template is identified by an ID number for faster analysis and efficient data storage. 

A \textit{log sequence} is a fragment of a log, i.e., a sequence of log messages contained in a log;  in some cases, it is convenient to abstract log sequences by replacing the log messages with their log templates. 
Log sequences are obtained by partitioning logs based on either log message identifiers (e.g., session IDs) or log timestamps (e.g., by extracting consecutive log messages using a fixed/sliding window). 
For a log sequence $l$, $|l|$ indicates the length of the log sequence, i.e., the number of elements (either log templates or log messages), not necessarily unique, inside the sequence.  

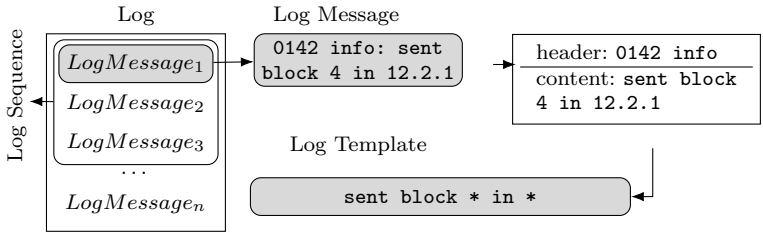
\begin{figure}
    \centering
\begin{tikzpicture}[
    node distance = 5mm and 7mm,
    module/.style={%
        draw, rounded corners,
        minimum width=#1,
        minimum height=5mm,
        font=\linespread{1}\selectfont
        },
    module/.default=2cm,
    >=LaTeX,
 disc/.style = {shape=cylinder, draw, shape aspect=0.27,
                shape border rotate=90,
                text width=20mm, align=center, font=\linespread{1}\selectfont},
  mdl/.style = {shape=ellipse, aspect=3, draw},
  alg/.style = {draw, align=center, font=\linespread{1}\selectfont},
  alg2/.style = {draw, align=center, 
  minimum width=#1,
        minimum height=5mm,
        font=\linespread{1}\selectfont},
        alg2/.default=2cm
                    ]
    \node [module = 2cm, fill= lightgray!60!white] (n1) {\small $LogMessage_1$};
    \node [below = 0.5mm of n1] (n2) {\small $LogMessage_2$};
    \node [below = 0.5mm of n2] (n3) {\small $LogMessage_3$};
    \node [below = 0.5mm of n3] (n4) {\small \dots};
    \node [below = 0.5mm of n4] (n44) {\small $LogMessage_n$};
    \node [fit=(n1)(n2)(n3), draw,rounded corners, inner
    sep=0.6mm] (fit0) {};
    \node [fit=(n1)(n2)(n3)(n4)(n44), draw, inner
    sep=1.4mm] (fit1) {};
    \node [above=0.015 mm of fit1] (l1) {Log};
    \node [above left=3.5 mm of fit0, rotate= 90] (n8) {Log Sequence};
    
    \node [alg2 = 2 cm, below right= 0.01 mm and 12 mm of l1, rounded corners, fill= lightgray!60!white](n5) { \small \texttt{0142 info: sent} \\ \small \texttt{block 4 in 12.2.1}};
    
    \node [alg2 = 5 cm, below right= 19.1 mm and 11.5 mm of l1, rounded corners, fill= lightgray!60!white](n7) {\small \texttt{sent block * in *}};
    
    \node [right = 13.5 mm of l1] (l2) {Log Message};
    
    \node [below=5.5 mm of n5] (l3) {Log Template};
    \node [alg2 = 1.5 cm, below right= 0.1 mm and 46 mm of l1](n6) {\begin{tabular}{l}
  header: \small \texttt{0142 info} \\
  \hline 
  content: \small \texttt{sent block} \\ \small \texttt{4 in 12.2.1}\\
  \end{tabular}};

    \draw [->] (n1) --(n5);
    \draw [->] (fit0) -- (-1.4,-0.5);
    \draw [->] (4.7, -0.01) -- (5, -0.01);
    \draw [->] (6.8, -1.12) |- (n7);
    \end{tikzpicture}
    \caption{An example illustrating the concepts of log, log message, log template, and log sequence}
    \label{fig:logs}
\end{figure}

\figurename~\ref{fig:logs} shows an example summarizing the aforementioned concepts. On the left side, the first three log messages are partitioned (using a fixed window of size three) to create a log sequence. The first message in the log sequence (\emph{LogMessage1}) is \texttt{0142 info: sent block 4 in 12.2.1}. It is decomposed into the header \texttt{0142 info} and the content \texttt{sent block 4 in 12.2.1}. The log template for the content is \texttt{sent block * in *}; the dynamic parts are \texttt{4} and \texttt{12.2.1}.

\subsection{Log Analysis Tasks}\label{sec:failurevsanomaly}

In the area of log analysis, several major tasks for reliability engineering, such as anomaly detection, and failure prediction, have been automated~\cite{He2021}; we provide an overview of these tasks below.

\subsubsection{Anomaly Detection}\label{sec:background:AD}
Anomaly detection is the task of identifying anomalous patterns in log data that do not conform to expected system behaviours~\cite{He2021}, indicating possible errors, faults, or failures in software systems. 

To automate the task of anomaly detection, log data is often partitioned into smaller log sequences. 
This partitioning is typically based on log identifiers (e.g., \textit{session\_ID} or \textit{block\_ID}), which correlate log messages within a series of operations; alternatively, when log identifiers are not available, timestamp-based fixed/sliding windows are also used. 
\citet{10.1145/3510003.3510155} assessed the accuracy of anomaly detection models considering both timestamp-based partitioning (with different time periods) and log identifier partitioning; models achieved higher accuracy and exhibited robustness when using the latter.

Labelling of partitions is then required, each partition usually being labelled as an anomaly either when an error, unknown, or failure message appears in it or when the corresponding log identifier is marked as anomalous. Otherwise, it is labelled as normal. 

\textit{Failure Detection.}
Failure detection is a special type of anomaly detection that specifically identifies failures within logs~\cite{Bogatinovski2022FailureIF}, as compared in~\figurename~\ref{fig:loganalysistasks}. Similar to anomaly detection, log data is partitioned into sequences. The decision of whether a log should be tagged as anomalous or a failure depends on the system being analyzed. By definition, anomaly detection targets a wide scope of abnormal behaviours (which may or may not be a system failure) whereas failure detection focuses on system failures.

\subsubsection{Failure Prediction}\label{sec:background:FP}
Failure prediction attempts to proactively generate alerts \textit{before} the occurrence of failures, which often lead to unrecoverable outages~\cite{He2021}. In failure prediction, a log is partitioned similarly to previous tasks, often using a session-based log identifier. 

\begin{figure}
\centering
\includegraphics[width=\textwidth]{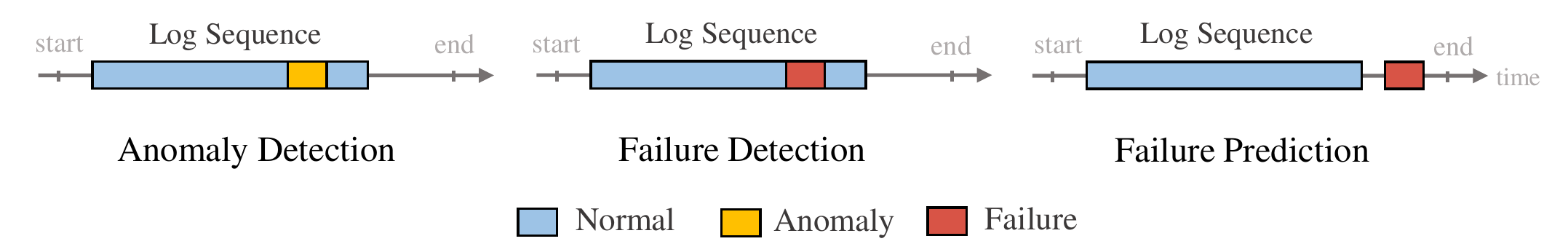}
\caption{Illustration of Log Analysis Tasks.}
\label{fig:loganalysistasks}
\end{figure}

The main differences between failure prediction and the above tasks are the following: 
\begin{itemize}
    \item \textit{mode of operation}. As shown in \figurename~\ref{fig:loganalysistasks}, anomaly or failure detection are reactive approaches that raise a flag once an anomaly or failure has happened. Instead, failure prediction is \textit{proactive}. It forecasts potential future failures, allowing enough time to address them. 
    \item \textit{input data}. The input of failure prediction typically consists of normal-looking inputs, a subset of which involves subtle and complex patterns in logs, which may be associated with a future failure. Patterns can indicate impending issues that have not yet manifested as failures in log data. 
\end{itemize}  

\begin{figure}
\centering
  \begin{subfigure}{.24\linewidth}
    \centering
    \includegraphics[width = \linewidth]{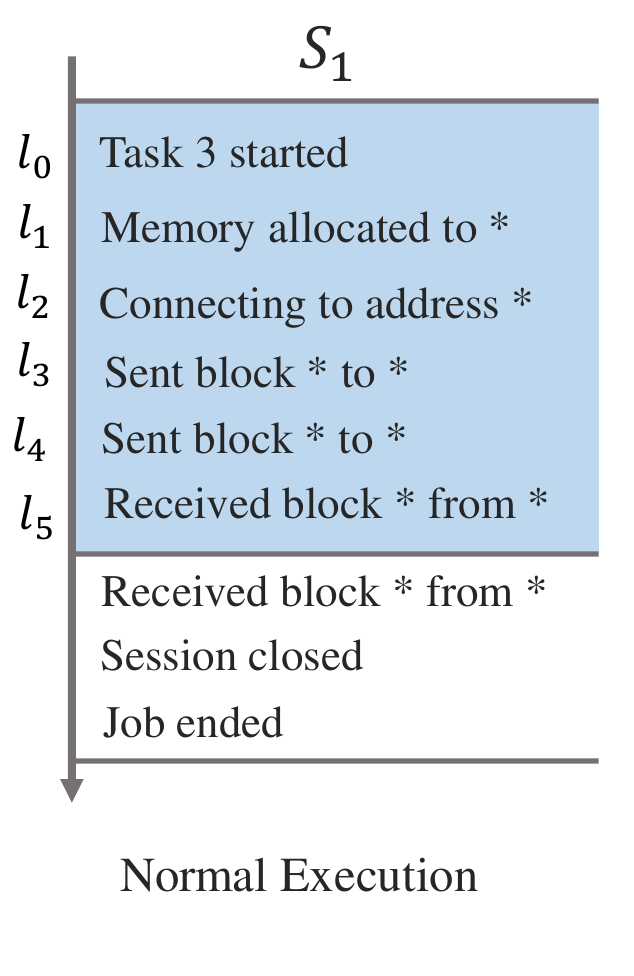}
    \caption{\label{fig:loganalysisexample-a}}
  \end{subfigure}%
  \hspace{0.3em}
  \begin{subfigure}{.24\linewidth}
    \centering
    \includegraphics[width = \linewidth]{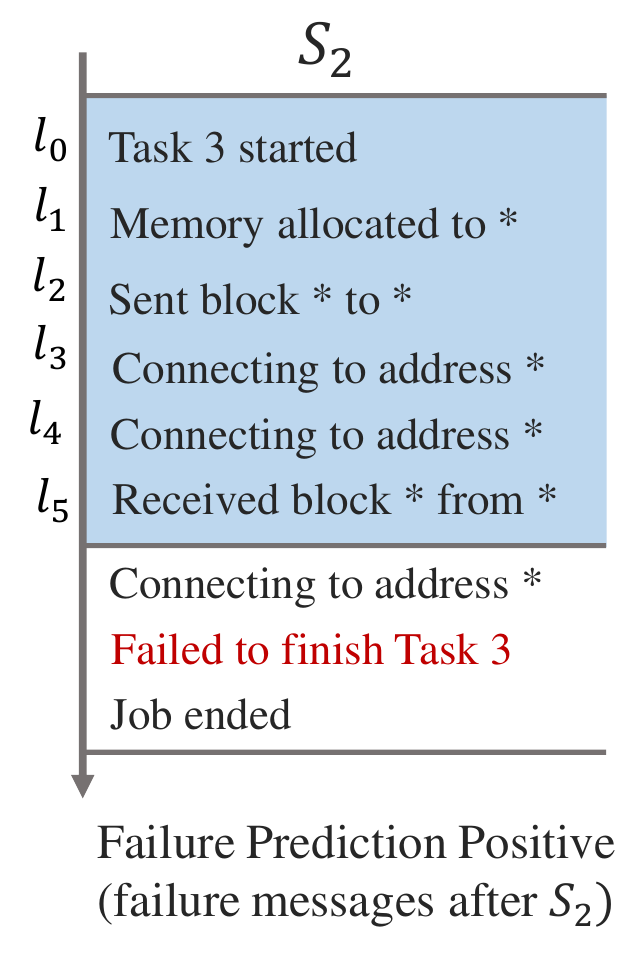}
    \caption{\label{fig:loganalysisexample-b}}
  \end{subfigure}%
  \hspace{0.3em}
  \begin{subfigure}{.24\linewidth}
    \centering
    \includegraphics[width = \linewidth]{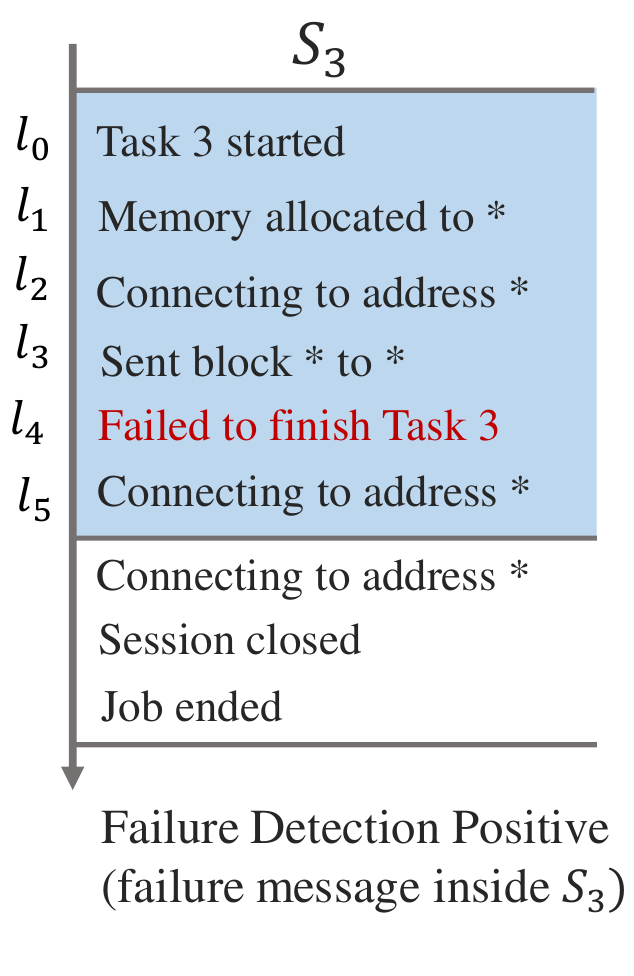}
    \caption{\label{fig:loganalysisexample-c}}
  \end{subfigure}
    \hspace{0.2em}
  \begin{subfigure}{.24\linewidth}
    \centering
    \includegraphics[width = \linewidth]{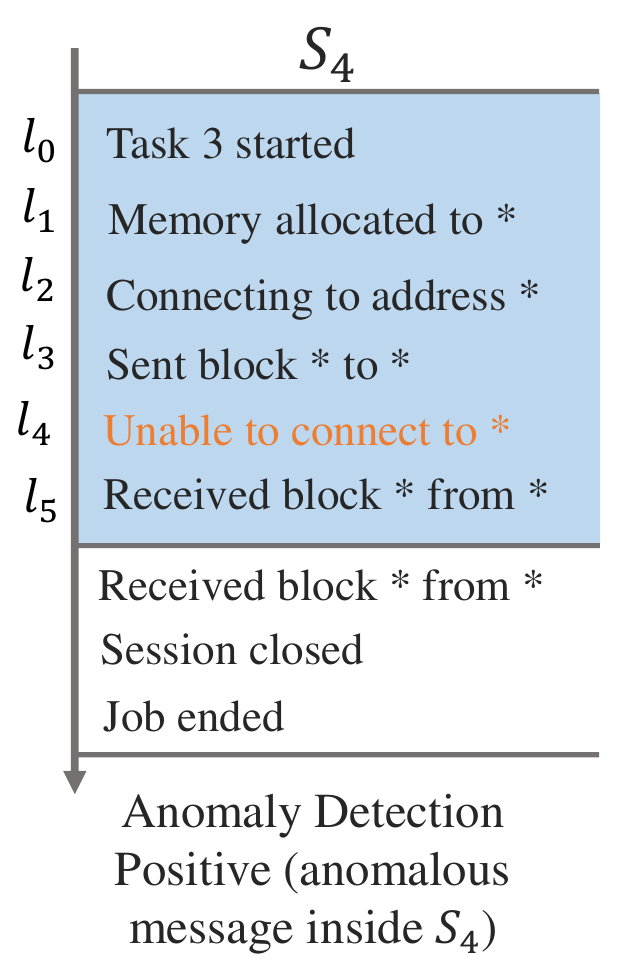}
    \caption{\label{fig:loganalysisexample-d}}
  \end{subfigure}
\caption{Comparison of Normal Sequence (on the left) and Positive Sequences in Log Analysis Tasks (on the right).\label{fig:loganalysisexample}}
\end{figure}

\figurename~\ref{fig:loganalysisexample} shows a simplified comparison of ``positive sequences'' (in contrast to ``normal'' sequences) for the aforementioned tasks (depicted in Subfigures~\ref{fig:loganalysisexample-b},~\ref{fig:loganalysisexample-c}, and~\ref{fig:loganalysisexample-d}), next to a normal log (depicted in Subfigure~\ref{fig:loganalysisexample-a}). The blue box in each Subfigure highlights a partitioned sequence of log templates, labelled as $S_1, S_2, S_3,$ and $S_4$. 
For failure prediction (see Subfigure~\ref{fig:loganalysisexample-b}), log templates in $S_2$ look normal when considered individually. However, their occurrence creates a pattern indicating a point on the timeline where a future failure, highlighted in red, happens. Hence, $S_2$ is a positive case in data labelling for failure prediction. 
Subfigure~\ref{fig:loganalysisexample-c}, on the other hand, shows $S_3$ as a positive instance for failure detection, since there is a failure message (also highlighted in red) within the blue sequence. Similarly, in anomaly detection, an anomalous log message, highlighted in yellow, appears within $S_4$ (see Subfigure~\ref{fig:loganalysisexample-d}). 

\textit{Dataset Transferability for Failure Prediction.}
It is worth mentioning that, as sketched in Subfigure~\ref{fig:loganalysisexample-d}, one cannot necessarily expect the occurrence of a failure after a log sequence with an anomalous section. That is, log data used for anomaly detection are not interchangeable with those intended for failure prediction. Therefore, using anomaly detection data for failure prediction would likely yield inaccurate and misleading results. 

When using data intended specifically for failure detection in the context of failure prediction, some assumptions should hold. First, log data is required to be ordered by timestamp, to make it possible to separate the sequence of messages before the occurrence of a failure. In addition, one must rigorously label log data to decide whether a sequence of log messages is indeed related to a future failure; this is especially challenging when there is no clear evidence of a failure, unlike many anomaly detection datasets. The quality of the initial labelling plays an important role as well. Considering the above, log data labelled for failure detection can be used for predictive tasks through careful preprocessing and rigorous validation (see \S~\ref{sec:real-world-dataset-processing}).

\subsection{DL Techniques in Log Analysis}\label{sec:background:ml-based-log-analysis}

In recent years, a variety of deep learning (DL) techniques have been applied to log analysis, and more specifically to failure prediction and anomaly detection. Compared to traditional ML techniques such as Random Forests (RF) and K-nearest Neighbours (KNN), DL techniques incrementally learn high-level features from data, removing complex feature extraction activities based on domain expertise. 

According to \citet{10.1145/3510003.3510155}, there are three main categories of DL approaches in log analysis: (1) Recurrent Neural network (RNN), (2) Convolutional Neural Network (CNN), and (3) transformer. Additionally, we have a new growing category called (4) Graph Neural Network (GNN). In each category, different variations can be adopted; for instance, Long Short-Term Memory networks (LSTM) and Bidirectional Long Short-Term Memory networks (BiLSTM), which fall into the RNN category, have been repeatedly used for anomaly detection and failure prediction~\cite{DeepLog, Das2018, LogRobust}. We now explain the major features of each category as well as their variations.

\subsubsection{RNN}
\label{sec:background:rnn} 

LSTM~\cite{HochSchm97, GersLSTM} is an RNN-based model commonly used in both anomaly detection and failure prediction~\cite{DeepLog,Das2018}. An LSTM network consists of multiple units, each of which is composed of a cell, an input gate, an output gate~\cite{HochSchm97}, and a forget gate~\cite{GersLSTM}. An LSTM-based model reads an input sequence $(x_1, \ldots, x_n)$ and produces a corresponding sequence $(y_1, \ldots, y_n)$ with the same length.
At each time step $t>1$, an LSTM unit reads the input $x_t$ as well as the previous hidden state $h_{t-1}$ and the previous memory $c_{t-1}$ to compute the hidden state $h_t$. The hidden state is employed to produce an output at each step.
The memory cell $c_t$ is updated at each time step $t$ by partially forgetting old, irrelevant information and accepting new input information. The forget gate $f_t$ is employed to control the amount of information to be removed from the previous context (i.e., $c_{t-1}$) in the memory cell $c_t$. 

As a recurrent network, an LSTM shares the same parameters across all steps, which reduces the total number of parameters to learn. Learning is achieved by minimizing the error between the actual output and the predicted output. Moreover, to improve the regularization of an LSTM-based model, a dropout layer is applied between LSTM layers. It randomly drops some connections between memory cells by masking their value.
LSTM-based models have shown significant performance in several studies in log-based failure prediction and anomaly detection ~\cite{LogAnomaly, Log2Vec, aarohi, Das2018}.

BiLSTM is an extension of the traditional LSTM~\cite{BiLSTMfirstpaper}. However, BiLSTM reads the sequence in both directions, enabling it to comprehend the relationships between the previous and the upcoming inputs. To make this possible, a BiLSTM network is composed of two layers of LSTM nodes, whereby each of these layers learns from the input sequence in the opposite direction. At time step $t$, the output $h_t$ is calculated by concatenating $h_t^f$ (the hidden states in a forward pass) and $h_t^b$ (the hidden states in a backward pass). By allowing this bi-directional computation, BiLSTM is able to capture complex dependencies and produce more accurate predictions. The BiLSTM-based model has achieved accurate results for anomaly detection~\cite{LogRobust}.

\subsubsection{CNN}
\label{sec:background:cnn} 

CNN is a neural network primarily employed for image recognition~\cite{CNNfirstpaper}. It has a unique architecture designed to handle 2D and 3D input data such as images and matrices. A CNN leverages convolutional layers to perform feature extraction and pooling layers to downsample the input. 

The 1D convolutional layer uses a set of filters to perform convolution operation with the 2D input data to produce a set of feature maps (CNN layer output). According to \citet{CNNdefinition}, let $w \in \mathit{R}^{k \times d}$ be a filter which is applied to a window of $k$ elements in a $d$-dimension input log sequence, and let $x_i$ represent the $i$-th elements in the sequence. A feature $c_i \in \mathit{R}$ is calculated as
$c_i = \sigma(w . x_{i:i+k-1}+b)$,
where $\sigma$ is the activation function (i.e., $ReLu$), $x_{i:i+k-1}$ represents the concatenation of elements $\{x_i, x_{i+1}, ..., x_{i+k-1}\}$, and $b \in \mathit{R}$ denotes a bias term. After this filter is applied to each window in the sequence ($\{x_{1:k}, x_{2:k}, ..., x_{n-k+1:n}\}$), a feature map $c = [c_1, c_3, ..., c_{n-k+1}]$ is produced, 
where $c \in \mathit{R}^{n-k+1}$. 
Parameter $k$ represents the kernel size; it is as an important parameter of the operation. Note that there is no padding added to the input sequence, leading to feature maps smaller than the input sequence. Padding is a technique employed to add zeros to the beginning and/or end of the sequence; it allows for more space for the filter to cover, controlling the size of output feature maps. Padding is commonly used so that 
the output feature map has the same length as the input sequence~\cite{padding}.

The pooling layer reduces the spatial dimensions of the feature maps extracted by the convolutional layer and simplifies the computational complexity of the network. 

Recently, CNNs have shown high-accuracy performance in anomaly detection~\cite{CNN}.

\subsubsection{Transformer}
\label{sec:background:transformer} 

The transformer is a type of neural network architecture designed for natural language processing tasks, introduced by \citet{AttentionIA}. The main innovation of transformers is the self-attention mechanism. More important parts of the input receive higher attention, which facilitates learning the contextual relationships from input data. This is implemented by calculating a weight for each input element, which represents the importance of that element with respect to the adjacent elements. Hence, a model with self-attention (not necessarily a transformer) can capture long-range dependencies in the input. 
Since the transformers do not process inputs sequentially like LSTM, positional encoding is needed. Positional encoding vectors are fixed-size, added to the input to provide information about the position of each element in the input sequence. Further, a transformer involves a stack of multiple transformer blocks. Each block contains a self-attention layer and a feed-forward neural network layer. In the self-attention layer, the model computes attention scores (weights) for each element, allowing it to capture the relationship between all input elements. 
The feed-forward layer is used to transform the representation learned by the self-attention layer into a new representation entering the next transformer block. In the area of log analysis, transformers have been recently applied in a few studies on anomaly detection~\cite{Neurallog, HitAnomaly, logBERT, LogSy},  showing outstanding performance. 

\subsubsection{GNN}
A Graph Neural Network (GNN) is a neural network designed to process data structured as graphs~\cite{GNN}. During training, taking a graph-structured input, it updates node feature vectors (where nodes are equivalent to vertices in the graph) iteratively with respect to feature vectors of its neighbour nodes and itself. By using the final feature vectors, GNNs can discern intricate relationships within the graph data. Hence, GNNs can be used for classification tasks at either the graph or node level.

The main difference between GNN and the aforementioned DL techniques is the data structure they process. GNNs process data structured as graphs. Since log data are initially sequential data, it requires further processing to construct a graph from sequential data. When a node represents a log template and a graph corresponds to a log sequence, classification at the graph level requires an aggregation method such as READOUT~\cite{GNN} to combine node feature vectors. 

We note that GNNs are sometimes regarded as a representation method, such as log sequence embedding strategies detailed in \S~\ref{sec:background:embed}, since they compute the graph representation of log sequences~\cite{Wu2023}. However, we consider them as a classification method, like the other DL methods described in \S~\ref{sec:background:ml-based-log-analysis}, since current GNNs necessitate pre-existing semantic embeddings for input.

\subsection{Log Sequence Embedding Strategies}\label{sec:background:embed}
When analyzing log sequences, the textual data of log sequences' elements must be converted into a vector representation that is understandable by a machine; such a conversion is called the \emph{log sequence embedding}. Generally, there are three main approaches for doing this: 
(1) template ID-based strategies such as count vectors~\cite{countvector}, (2) semantic-based strategies based on the contextual information of sequence elements, or (3) hybrid strategy as a combination of the previous two strategies.
Here, we cover one widely used example for each case in the following sections.

\subsubsection{Template ID-based Strategy}\label{sec:background:logkey2vec}
There are many studies that have achieved high accuracy results by using log embedding strategies that rely on the ID numbers or count vectors of log sequence elements~\cite{experimentreport}. Advantages include the speed of processing and model simplicity since text preprocessing (e.g., tokenization) is not required. 
However, they do not consider the order of log messages (templates) in a log sequence, making them prone to unreliable results when the sequential pattern of log messages (templates) matters (e.g., in failure prediction). 

TF-IDF~\cite{Rajaraman2011} is a widely used embedding strategy in data mining and information retrieval, employed for log analysis at two different levels: log template ID level and word (token) level. 
At the log template ID level, it measures the frequency of a unique log template in a log sequence, term frequency (TF), divided by how common this log template is in the total dataset (i.e., Inverse Document Frequency - IDF). 
At the word (token) level, it delves deeper. It calculates the TF-IDF value for each unique word (token) inside a log template and assigns the aggregated value to a log template. Both TF-IDFs compute an embedding vector for each log sequence, making it incompatible with methods requiring an embedding vector at the log template level.

Logkey2vec, introduced by~\citet{CNN}, is another strategy used in log analysis, which is based on log template IDs and is able to transform a log template into an embedding vector.
Logkey2vec maps each unique log template ID to a vector representation. 
It is a trainable layer implemented inside a neural network. It relies on a matrix called ``codebook'', where the number of rows is the vocabulary size and the number of columns is the embedding vector size of each log template ID. The embedding vectors are first initialised by random numbers and are improved through backpropagation during training. For a log sequence, Logkey2vec computes the embedding vector of each log template based on its log template ID; each row of the matrix represents the whole log sequence. 
We note that Logkey2vec is not semantic-based in a linguistics sense since it solely takes log template IDs as input, disregarding the semantic information that lies in the text of log templates. Moreover, unlike tools such as word2vec~\cite{word2vec}, which is pre-trained using CBOW (Continuous Bag-of-Words) and Skip-grams~\cite{word2vec}, Logkey2vec is not pre-trained by any method; it requires the aforementioned training on its target log data. This strategy has also been applied, with a different name, by~\citet{Bogatinovski2022FailureIF} (who used the term ``vectorizer''), and by~\cite{logBERT} (who used the term ``Embedding Matrix'').

\subsubsection{Semantic-based Strategy}\label{sec:background:bert}
Studies using semantic-based strategies take into account the linguistic relationship between words in log templates. In 2019, \citet{LogAnomaly} proposed \textit{template2vec}, an embedding strategy based on synonyms and antonyms relation of words mentioned in log data. This strategy enables the matching of new log templates with existing ones. However, since it is trained on manually added domain-specific synonyms and antonyms, its applicability is limited.

In the past few years, Bidirectional Encoder Representations from Transformers (BERT) has provided significant improvements in the semantic embedding of textual information by taking the contextual information of text into account. It has been used in a few studies in log sequence embedding~\cite{logBERT, Neurallog}. This model fares better than the other pretrained transformer-based models: GPT2~\cite{GPT2} and RoBERTa~\cite{Roberta} in log sequence embedding~\cite{Neurallog}.

The pre-trained BERT base model~\cite{BERTbasemodel} provides the embedding matrix of log sequences where each row is the representation vector of its corresponding log template inside the sequence. The BERT model is applied to each log template separately and then the representation is aggregated inside a matrix. To embed the information of a log template into a 768-sized vector, the BERT model first tokenizes the log template text. BERT tokenizer uses WordPiece~\cite{wordpiece}, which is able to handle out-of-vocabulary (OOV) words to reduce the vocabulary size. Further, the tokens are fed to the 12 layers of BERT's transformer encoder. After obtaining the output vectors of a log template's tokens, the log template embedding is calculated by getting the average of output vectors. This process is repeated for all the log templates inside the log sequence to create an $n \times 768$ matrix representation where $n$ is the size of the log sequence.

\subsubsection{Hybrid Strategy}\label{sec:background:hybrid}
This category aims to combine the benefits of both template ID-based and semantic-based strategies while compensating for their limitations. The main instance of this category is the study of~\citet{LogRobust}. They leverage FastText~\cite{fasttext} to convert each word (or token) of a log template into a $d$-dimensional vector ($d=300$). FastText is a word vectorisation tool pre-trained on the Common Crawl Corpus dataset~\cite{commoncrawl}; it converts words into vectors while capturing their semantic relationship. Consequently, words having similar meanings result in similar vectors. The word vectors are further aggregated into one vector representing a log template using a weighted average with TF-IDF (calculated at the word level). Specifically, consider a log template $T$ consisting of a list of words, $[t_1, t_2,...,t_N]$, where $N$ indicates the number of words. The list of words can be represented as a list of vectors $[v_1, v_2,...,v_N]$, where $v_i \in \mathit{R}^d$ is a semantic vector of $t_i$. The embedding vector of $T$, $V_T$, is then calculated according to Equation~\ref{eq:TF-IDFaggression}, where $w_i \in \mathit{R}$ indicates the TF-IDF value of $t_i$.
\begin{equation}
    V_T = \frac{1}{N}\sum_{i=1}^{N}w_i.v_i
    \label{eq:TF-IDFaggression}
\end{equation}
This strategy seeks to retain the advantages of the previous strategies. If a word is frequently mentioned among log templates, it is given a  lower TF-IDF weight during the aggregation of word vectors, increasing the distinction of embedding vectors between log templates. Moreover, similar to BERT but not as informative in terms of word context, FastText assigns vectors with high cosine similarity to two log templates that contain different words but are semantically close.
\section{Related Work}\label{sec:related-works}

 In this section, we will first discuss empirical studies on log-based anomaly detection and move on to more closely related failure prediction studies. We will also discuss studies related to dataset synthesis at the end.

\subsection{Related Empirical Studies}

\subsubsection{Log-based Anomaly Detection}\label{sec:related-works:AD}
As discussed in \S~\ref{sec:failurevsanomaly}, anomaly detection is a different task than failure prediction. However, since they are both binary classification tasks on log data, they can rely on similar DL architectures~\cite{Das2018, aarohi}.
There are several papers reporting empirical studies of different DL-based methods for log-based anomaly detection.
Due to the large number of works and differences in objectives, in our review, we include studies that covered more than one DL model, possibly based on the same DL-based approaches.

Table~\ref{tab:relatedworks} briefly summarises anomaly detection studies including empirical evaluations. 
Column ``DL Type(s)'' indicates the type of DL network covered in each paper. We indicate the Log Sequence Embedding (LSE) strategies, introduced in \S~\ref{sec:background:embed}, in the next column; notice there are a few models not using LSE, such as DeepLog~\cite{DeepLog}. 
Column ``Dataset(s)'' indicates which datasets (whether existing datasets or synthesised ones) were used in the studies.
 Column ``Control of Dataset Characteristics'' indicates whether the dataset characteristics were controlled during the experiment and lists such characteristics. 
 In the last column, the labelling scheme indicates the applied method(s) for log partitioning, as mentioned in \S~\ref{sec:background:logs}, based either on a log identifier or on timestamp (represented by \textit{L} and \textit{T}, respectively).

\begin{table*}[htbp]
    \caption{Overview of Related Empirical Studies}\label{tab:relatedworks} 
    \centering
    \begin{threeparttable}
\resizebox{\columnwidth}{!}{
    \begin{tabular}{cccccc} 
        \toprule
        \textbf{\small Paper} & \textbf{\small DL Type(s)} & \textbf{\small LSE Strategi(es)} & \textbf{\small Dataset(s)} & \textbf{\small \makecell{Control of\\ Dataset\\ Characteristics}} & \textbf{\makecell{\small Labelling \\Scheme}}\\
        \midrule
        \addlinespace

        \multicolumn{6}{c}{\textbf{Anomaly Detection}}\\
        \midrule
        \addlinespace
        
        \citet{CNN} & \centering LSTM, CNN, MLP & \centering Logkey2vec & \centering HDFS & \centering No & L\\[0.2cm]
        \addlinespace
        
        \makecell{Meng et \\al.\cite{LogAnomaly}} & \centering LSTM & \centering \makecell{template ID,\\ Template2Vec} & \centering HDFS, BGL & \centering No & T, L\\[0.2cm]
        \addlinespace
        
        \makecell{Huang et\\ al. \cite{HitAnomaly}} & \centering \makecell{LSTM, BiLSTM, \\Transformer} & \centering \makecell{count vector,\\ F+T, \\Log Encoder} & \centering \makecell{HDFS, BGL, \\OpenStack} & \centering \makecell{Yes (unstable log \\injection ratio)} & T, L\\[0.2cm]
        \addlinespace
        
        \makecell{Yang et \\al. \cite{PLELog}} & \centering \makecell{LSTM, BiLSTM, \\GRU} & \centering \makecell{template ID, \\TF-IDF, F+T} & \centering HDFS, BGL & \centering No & T, L\\[0.2cm]
        \addlinespace
        
        \makecell{ Guo et al. \\\cite{logBERT}} & \centering LSTM, Transformer & \centering \makecell{template ID,\\ count vector, \\Embedding \\Matrix} & \centering \makecell{HDFS, BGL, \\Thunderbird} & \centering No & T, L\\[0.2cm]
        \addlinespace
        
        \makecell{Le and Zhang\\ 2021 \cite{Neurallog}} & \centering \makecell{LSTM, BiLSTM, \\Transformer}& \centering \makecell{count vector,\\Log2Vec*, \\  F+T, BERT}&\centering \makecell{HDFS, BGL, \\Spirit, \\Thunderbird} & \centering No &T, L\\[0.2cm]
        \addlinespace

         \makecell{ Bogatinovski\\ et al. \cite{Bogatinovski2022FailureIF}} &  \centering \makecell{LSTM, \\Transformer} &  \centering \makecell{count vector, \\vectorizer} &  \centering \makecell{OpenStack\_v2} &  \centering \makecell{Yes (unstable log \\injection ratio)} &  T\\[0.2cm]
        \addlinespace
        
        \makecell{ Le and Zhang\\ 2022 \cite{10.1145/3510003.3510155}} & \centering \makecell{LSTM, BiLSTM, \\GRU, CNN} & \centering \makecell{template ID, \\Logkey2vec, \\F+T}& \centering \makecell{HDFS, BGL,\\ Spirit, \\Thunderbird} & \centering \makecell{Yes (class \\distribution, data\\ noise, partitioning\\ methods)} &T, L\\[0.2cm]
        \addlinespace

         \makecell{ \citet{logGD}} &  \centering \makecell{BiLSTM, CNN\\Transformer, \\GNN} &  \centering \makecell{count vector, \\Logkey2vec,\\ F+T, BERT}&  \centering \makecell{HDFS, BGL,\\ Spirit, \\Thunderbird} &  \centering \makecell{Yes \\(partitioning\\ methods)} & T, L\\[0.2cm]
        \addlinespace

         \citet{Wu2023} &  \centering \makecell{MLP, CNN \\LSTM} &  \centering \makecell{count vector,\\ TF-IDF, \\ Word2Vec, \\FastText, BERT\\} &  \centering \makecell{HDFS, BGL,\\ Spirit, \\Thunderbird} &  \centering \makecell{Yes \\(partitioning\\ methods)} &  T, L\\
        \midrule
        \addlinespace

        \multicolumn{6}{c}{\textbf{Failure Prediction}}\\
        \midrule
        \addlinespace

         \citet{unbalance_failure} &  \centering BiLSTM &  \centering N/A &  \centering AzureML &  \centering No &  T\\[0.2cm]
        \addlinespace

         \makecell{Das et al.\\2018~\cite{Das2018}\\2020~\cite{aarohi}} &  \centering LSTM &  \centering template ID &  \centering  Clay-HPC &  \centering No &  L\\[0.2cm]
        \addlinespace
        
        Our Study**  & \centering \makecell{LSTM, BiLSTM, \\CNN, \\ Transformer} & \centering \makecell{Logkey2vec,\\ BERT,  F+T} & \centering \makecell{Synthesized \\Data, \\ OpenStack\_FP }& \centering \makecell{Yes (Dataset size,\\ Failure Percentage,\\ LSL, Failure \\Pattern type)} & L\\
        \midrule
        
    \end{tabular}
    }

\begin{tablenotes}\scriptsize\item[]
      \textit{*: we highlight that Log2Vec is different than Logkey2vec, a log sequence embedding strategy \\(see \S~\ref{sec:background:logkey2vec})}
      \textit{**: further discussed in \S~\ref{sec:discussion}}
\end{tablenotes}

\end{threeparttable}
\end{table*}

We now briefly explain the included papers with the aim of motivating our study and highlighting the differences. We note that, unless we mention it, LSE strategies are implemented specifically for one DL model (combinations are not explored). Indeed, many of the reported techniques tend to investigate one such embedding strategy or simply do not rely on any. The studies are listed in chronological order. 
\citet{CNN} (2018) introduced CNN for anomaly detection as well as the Logkey2vec embedding strategy (see \S~\ref{sec:background:logkey2vec}). They compared it to LSTM and MLP networks, also relying on the Logkey2vec embedding strategy.
\citet{LogAnomaly} (2019) developed LogAnomaly, an LSTM-based model, using their proposed embedding strategy, Template2Vec (a log-specific variant of Word2Vec). 

The first study considering transformers in their DL comparison is by \citet{HitAnomaly} (2020), featuring three DL models: HitAnomaly (transformer-based), LogRobust~\cite{LogRobust} (BiLSTM-based), and DeepLog (LSTM-based). HitAnomaly utilises transformer blocks (see \S~\ref{sec:background:transformer}) as part of its LSE strategy, called Log Encoder. LogRobust employed the hybrid strategy of FastText and TF-IDF shows as F+T while DeepLog did not utilise any LSE strategy. The authors also controlled dataset characteristics by manipulating the unstable log ratios. 
\citet{PLELog} (2021) proposed the GRU-based~\cite{cho2014learning} PLELog and compared it to LogRobust and DeepLog. PLELog used the TF-IDF technique and LogRobust used F+T.
\citet{logBERT} (2021) proposed a transformer-based model, LogBERT, and compared its performance with two LSTM-based models, LogAnomaly and DeepLog. LogBERT uses an Embedding Matrix for its embedding strategy, which is similar to Logkey2vec.
\citet{Neurallog} (2021) evaluated their proposed transformer-based model, Neurallog, against LogRobust (BiLSTM-base) and DeepLog (LSTM-based). The LSE strategies for the models were a pre-trained BERT (see \S~\ref{sec:background:bert}) for Neurallog and Log2Vec~\cite{Log2Vec} (a strategy based on Word2Vec) for DeepLog.

An important recent work on failure detection is the study of~\citet{Bogatinovski2022FailureIF}. They presented log data as sequences of subprocesses instead
of sequences of log templates. To this end, they used transformer-based network and clustering methods to extract subprocesses and further leverage them to detect failure using an HMM~\cite{HMM}. For LSE, they designed the ``vectorizer'' that is similar to Logkey2vec. Their work includes the evaluation of varying unstable log ratios and their impact on their model performance.

\citet{10.1145/3510003.3510155} (2022) conducted a comprehensive evaluation of several DL models including LSTM-based models such as DeepLog and LogAnomaly, GRU-based model PLELog, BiLSTM-based model LogRobust, and CNN. The study focused on various aspects including data selection, data partitioning, class distribution, data noise, and early detection ability.
 Although they provide insights on many models and dataset characteristics, they did not include transformer-based models such as Neurallog, or recent semantic-based LSE strategies like BERT, and are limited to commonly used datasets.

\citet{logGD} (2022) proposed a GNN-based anomaly detection model, LogGD, and compared it with DL-based models from three categories: CNN, LogRobust (which is BiLSTM-based), and NeuralLog (which is transformer-based). Both NeuralLog and LogGD leverage BERT to extract semantic embeddings from log sequences. While \citet{logGD} took into account a wide range of DL techniques and LSE strategies from each category, their results, similar to those of \citet{10.1145/3510003.3510155} (2022), were obtained using only public datasets. 

Finally, \citet{Wu2023} (2023) studied the effectiveness of different LSE on ML-based models for anomaly detection. In contrast to the study of \citet{10.1145/3510003.3510155}, they explored all the possible combinations between LSE strategies and DL techniques and provided an accurate ranking for each category. They included six LSE strategies: Count Vector and TF-IDF (the word-level and template-level) as template ID-based strategies, and Word2Vec, FastText, and BERT as semantic-based strategies. However, they did not consider hybrid strategies. DL techniques are limited to MLP, CNN, and LSTM, while the rest of the common methods such as BiLSTM and transformers are left out. Similarly to \citet{10.1145/3510003.3510155}, their results are bound to four public datasets.

\paragraph{Datasets.} 
Studies relying on publicly available datasets are limited to the following: Hadoop Distributed File System (HDFS) collected in 2009, and three HPC datasets, BGL, Spirit, and Thunderbird, collected between 2004 and 2006. 
Besides, for failure detection, there is the OpenStack dataset (2017) created by injecting a limited number of bugs at different execution points. In 2022, thanks to the effort of~\citet{Bogatinovski2022FailureIF}, OpenStack was labelled at the log message level, which we refer to as OpenStack\_v2. Overall, due to the limited number of available public datasets, there is a growing number of works focusing either on labelling existing data to a deeper level or on synthesising log data, as discussed in the following section.

\subsubsection{Log-based Failure Prediction}\label{sec:related-works:FP}
In recent years, there have been a number of studies on log-based failure prediction, especially in large-scale systems where signs of failure may not be obvious. 
Early works on failure prediction focused on structured logs (e.g., numeric parameters) mined from system logs. \citet{Sahoo2003} collected system health status logs and employed several time-series models such as the mean of previous values to predict indicative metrics (e.g., system utilization percentage, network IO usage, and system idle time). \citet{Russo2015} applied different SVMs relying on radial basis function and linear kernels that take multi-dimensional data representing values for each of the metrics to predict a future log sequence related to a failure.  More recently, \citet{unbalance_failure} proposed a method that combines two ML models, BiLSTM and RF, to process temporal and 
 spatial data, respectively, and concatenates their outputs to predict the likelihood of a node failing in the near future.
 \citet{prefix} expanded this task to semi-structured logs. They extracted log templates from raw syslog messages and derived features from sequences of log templates. By training an RF-based model, Prefix, on features of previously seen log datasets, they achieved high accuracy in switch failure prediction compared to SVM and HMM. More recently, \citet{rfvsxgboost}  adopted machine learning models to predict system crashes on cloud service data; in their study, RF achieved the best accuracy compared to xgboost and SVM.
The study of \citet{Das2018} opened the door to analyzing semi-structured logs using DL. After extracting unique log templates, they derived patterns from them leading to a failure using LSTM. Following that, in 2020, they introduced an improved LSTM-based model, Aarohi~\cite{aarohi}, as state-of-the-art with faster inference time. Both Dash and Aarohi rely on the template ID-based strategy for embedding (see \S~\ref{sec:background:embed}).  The above DL-based studies of failure prediction are briefly summarised in Table~\ref{tab:relatedworks}.

\paragraph{Datasets.} Due to security concerns, in many reported works in the literature, the data sources are unavailable including the Clay-HPC (Clay high-performance computing (HPC) systems) dataset applied on Aarohi~\cite{aarohi} and Dash~\cite{Das2018}.
On the other hand, the Prefix dataset is available but is of limited use, due to its low complexity leading easily to high accuracy regardless of the approach.
As a result, 
We found the limited number of publicity available datasets to be a hindrance. We therefore opted to develop a method for synthesising new datasets, as described next.

\subsection{Dataset Synthesis Algorithms}\label{sec:rw:data-syn}
In the log analysis literature, especially in anomaly detection, dataset synthesis refers to the modification of an existing dataset to simulate specific scenarios, such as system performance issues~\cite{swisslog}, or evaluation of logs driven from system updates~\cite{LogRobust, HitAnomaly}.
On the other hand, in closely related literature on system monitoring, there are data synthesis algorithms for trace and benchmark generation that can create new data without relying on an existing dataset~\cite{Blom2005, Bombarda2020}.
Given the restrictions of available and suitable datasets for our failure prediction problem (as discussed in \S~\ref{sec:failurevsanomaly}), we henceforth refer to the second group of algorithms when mentioning data synthesis.

In 2005, \citet{Blom2005} proposed a method for generating test suites for systems whose behaviours can be described by extended finite state machines (EFSM). This method produces a test sequence, referred to as a trace, that represents a coverage item. An \textit{observer} monitors the trace and ``accepts" it in case the specified coverage item has been covered. More recently, in 2017, \citet{Kluge2017} introduced EMSBench, which contains a model capable of mimicking complex system behaviour. Using this model, sequential traces are generated for the purpose of comparing different platforms. In 2020, \citet{Bombarda2020} leveraged FSM to design an algorithm that produces test sequences in the form of traces, identifying those with invalid inputs. By employing FSM, they successfully embedded the system constraints into the FSM during the generation process, ensuring the creation of only valid test sequences. Furthermore, \citet{Krstic2020} presented an algorithm for generating an event stream with their associated arbitrary values. These logs are compatible with system specifications in the first-order dynamic logic (MFODL)~\cite{MFOCL}. 
We will further discuss existing data synthesis algorithms and their differences with ours when we present the latter in \S~\ref{sec:syn-data-gen}.

\section{Failure Prediction Architecture}\label{sec:architecture}

This section introduces our modular architecture for failure prediction, which aims to help us systematically evaluate various embedding strategies and DL encoders. Moreover, this modular architecture can serve as a baseline architecture for follow-up studies. Therefore, we describe it in this section, independently from the description of the empirical study design (see Section~\ref{sec:empirical-design}).

\begin{figure}
\centering
\begin{tikzpicture}
[
    node distance = 5mm and 7mm,
    module/.style={%
        draw, rounded corners,
        minimum width=#1,
        minimum height=5mm,
        font=\linespread{1}\selectfont
        },
    module/.default=2cm,
    >=LaTeX,
 disc/.style = {shape=cylinder, draw, shape aspect=0.27,
                shape border rotate=90,
                text width=20mm, align=center, font=\linespread{1}\selectfont},
  mdl/.style = {shape=ellipse, aspect=3, draw},
  alg/.style = {draw, align=center, font=\linespread{1}\selectfont},
  alg2/.style = {draw, align=center, 
  minimum width=#1,
        minimum height=5mm,
        font=\linespread{1}\selectfont},
        alg2/.default=2cm
                    ]
    \node [disc, fill= lightgray] (t1) {Log Data};
    
    \node [alg=2cm, fill = lightgray!30!white, draw = lightgray!30!white, below= 12.5mm of t1] (I1) {log sequence \\\small (\texttt{e.g.} $(x_1, x_2, ..., x_i,...,x_n)$\\ \small $x_i:$ \texttt{process * passed})};
    
    \node [above = 1 mm of I1] (o2) {\small Input Unit};
    
    \node [alg2=1.5cm,  right= 4 mm of I1, rounded corners] (I2) {Embedding \\technique\\\small(\texttt{e.g. BERT})};
    
    \node [alg=2cm, fill = lightgray!30!white, draw = lightgray!30!white, right= 4 mm of I2] (I3) {$n \times \theta$ matrix\\\small (\texttt{e.g.} \\\small$x_i:[0.56, ..., 0.87]$) };
    
    \node[above=2 mm of I2] (L1) {Embedding Step};
    
    \node[fit=(I1) (I2) (I3)(L1), draw, rounded corners, inner sep=1.4mm] (fit1) {};
    
    \node [alg2=2.6 cm, rounded corners, below=19 mm of I1] (m1) {DL encoder\\ \small \texttt{(e.g. CNN,RNN,}\\\small\texttt{ Trandsformer)}};
    
    \node [above = 1 mm of m1] (l1) {\small};
    
    \node [alg2=1 cm, fill = lightgray!30!white, draw = lightgray!30!white, right= 3 mm of m1] (o1) {feature\\vector\\of size $m$};
    
    \node [alg2=0.6cm, right = 5 mm of o1, rounded corners] (m2) {\small FFN};
    
    \node [alg2=1cm, fill = lightgray!30!white, draw = lightgray!30!white, below= 21 mm of I3] (m3) {normal or\\ failure?};
    
    \node [above = 1 mm of m3] (o3) {\small Output};
    
    \node[below=13 mm of I2] (L2) {Classification Step};
    
    \node[fit=(m1) (m2) (m3)(L2)(o1), rounded corners, draw, inner
    sep=1.4mm] (fit2) {};
    
    \draw[->] (t1)--(0,-0.8);
    \draw[->] (I1)--(I2);
    \draw[->] (I2)--(I3);
    \draw[-] (5.5, -3.25)|- (0, -3.8);
    \draw[->] (0, -3.8) -- (0, -4.25);
    \draw[->] (m1)--(o1);
    \draw [->] (o1)--(m2);
    \draw[->] (m2)--(m3);
\end{tikzpicture}
\caption{Overview of the modular architecture for failure  prediction \label{fig:base}}
\end{figure}
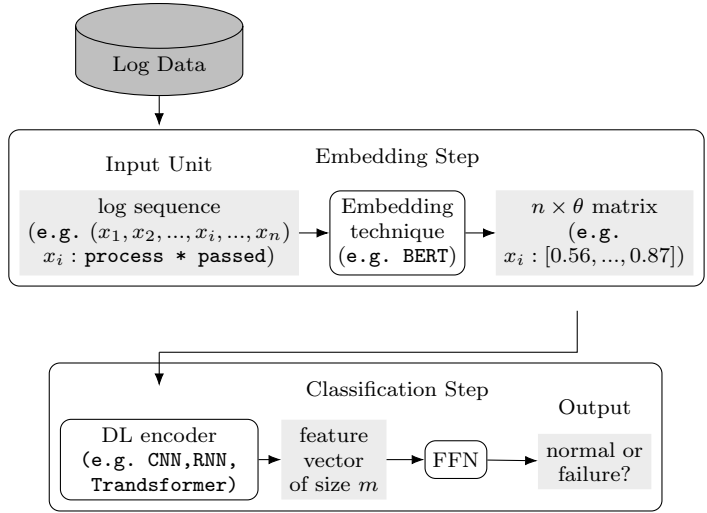

\figurename~\ref{fig:base} depicts the modular architecture. 
The architecture consists of two main steps, \textit{embedding} and \textit{classification}, allowing for different embeddings and DL techniques, respectively.
We note that preprocessing is not required in this architecture since log sequences are based on log templates which are already preprocessed from log messages. 

In the embedding step, log sequences are given as input, and each log sequence is in the form  $(x_1,x_2,...,x_i,..,x_n)$, where $x_i$ is a log template ID and $n$ is the length of the log sequence. 
An embedding technique (e.g., BERT) converts each $x_i$ to a $\theta$-dimensional vector representing the semantics of $x_i$, where $\theta$ is the size of log sequence embedding.
Then each log sequence forms a matrix $X \in \mathit{R}^{n\times\theta}$. Different log sequence embedding strategies can be applied; more information is provided in \S~\ref{sec:arch:LM}. 

In the classification step, the embedding matrix is processed to predict whether the given log sequence leads to a failure or not. 
A DL model, as an encoder $\Phi$ encodes the matrix $X$ into a feature vector $z = \Phi(X) \in \mathit{R}^{m}$, where $m$ is the number of features, which is a variable depending on the architecture of $\Phi$. 
Different DL encoders can be applied; more information is provided in \S~\ref{sec:arch:DL}. 
Similar to related studies~\cite{HitAnomaly, CNN}, the output feature vector $z$ is then fed to a feed-forward network (FFN) and softmax classifier to create a vector of size $d$ ($d=2$), capturing the prediction of the input unit label. As the FFN has a consistent setting across various configurations, it is separated as a common,  trainable part of the architecture, following an architecture similar to the one of the NeuralLog model~\citet{Neurallog} as well as the LogRobust one~\citet{LogRobust}.

More specifically, the FFN activation function is rectified linear unit (ReLu), and the output vector of the FNN $r$ is defined as $r = max(0, z W_1+b_1)$ where $W_1 \in \mathit{R}^{m\times d_{f}}$ and $b_1 \in \mathit{R}^{d_{f}}$ are a trainable parameter, and $d_{f}$ is the dimensionality of the FNN. 
Further, the calculation of the softmax classifier is as follows.
\begin{align}
    o = rW_2+b_2\\
    \mathit{softmax}(o_p) = \frac{exp(o_p)}{\sum_{j}{exp(o_j)}}
\end{align}
where $W_2\in \mathit{R}^{d_f \times d}$ and $b_2 \in \mathit{R}^{d}$ are trainable parameters to convert $r$ to $t \in \mathit{R}^{d}$ before applying softmax; $o_p$ represents the $p$-th component in the $o$ vector, and $exp$ is the exponential function. After obtaining the softmax values, the position with the highest value determines the label of the input log sequence.

Overall, the configuration of an embedding strategy and a DL encoder forms a language model that takes textual data as input and transforms it into a probability distribution~\cite{AttentionIA}. This language model handles the log templates as well as learning the language of failure patterns to predict the label of sequences. 

To train the above architecture, a number of hyper-parameters should be set such as the choice of the optimizer, loss function, learning rate, input size (for some deep learning models), batch size, and the number of epochs. Tuning these hyper-parameters is highly recommended as it significantly increases the chances of achieving the best failure prediction accuracy. Section~\ref{sec:training-testing} will detail the training and hyper-parameter tuning in our experiments.

After the model is trained, it is evaluated with a test log split from the dataset with stratified sampling. We used stratified sampling to keep the same proportion of failure log sequences as in the original dataset. Similar to training data, the embedding step transforms the test log sequences into embedding matrices. The matrices are then fed to the trained DL encoder to predict whether log sequences lead to failure or not.

\subsection{Embedding Strategies.}\label{sec:arch:LM}
While the modular architecture can accommodate various log sequence embedding options, we only consider one representative instance from each of three LSE strategies (see \S~\ref{sec:background:embed}), given our experimental constraint. More details are provided in \S~\ref{sec:design:DL&embed}.

Note that following three techniques were not compared in the same study before, according to Table~\ref{tab:relatedworks}. 

\paragraph{Logkey2vec.}
For Logkey2vec (see \S~\ref{sec:background:logkey2vec}), we set the embedding size to \num{768}, similar to BERT for better comparison. The vocabulary size is set to 200, consistent with the study of \citet{CNN}.

\paragraph{BERT.}
The maximum number of input tokens for BERT (see \S~\ref{sec:background:bert}) is \num{512} tokens. This limit does not constitute a problem in this work since the log templates in our datasets are relatively short and the total number of tokens in each log template is always less than \num{512}. Even if log templates were longer than 512, there are related studies suggesting approaches to use BERT accordingly~\cite{9587007, Ding2020CogLTXAB, 10.1007/978-3-030-32381-3_16}.
Each layer of the transformer encoder contains multi-head attention sub-layers and FFNs to compute a context-aware embedding vector ($\theta=768$) for each token.
This process is repeated for all the log templates inside the log sequence to create a matrix representation of size $n \times 768$, where $n$ is the length of the input log sequence.

\paragraph{FastText+TF-IDF.} Following its initial evaluation~\cite{LogRobust}, the dimension of the embedding vector is set to 300 ($d=300$).

\subsection{Deep Learning Encoder}\label{sec:arch:DL}
In this section, we illustrate the main features of the four DL encoders that can be used in the ``Classification step'' when instantiating our base architecture. We selected four encoders (LSTM-, BiLSTM-, CNN-, and transformer-based) because they cover the main DL types. 
 These four encoders cover all the DL techniques used in log-based failure prediction (BiLSTM and LSTM). Additionally, they represent the most common DL techniques used in relevant log analysis tasks: LSTM has been employed in nine studies, BiLSTM and transformers in five, and CNN in three, as detailed in Table~\ref{tab:relatedworks}.
GNNs are not included because there is no fair way to compare them with the others due to the required pre-processing stage required to transform sequential data into graphs, which is an expensive endeavour and a subject of current research~\cite{logGD}.

\paragraph{LSTM-based.}
This DL model is inspired by the LSTM architecture suggested by related works, including DeepLog~\cite{DeepLog}, Aarohi~\cite{aarohi}, and Dash~\cite{Das2018}. The model contains one LSTM hidden layer with \num{128} nodes and ReLu activation. A Dropout with a rate of \num{0.1} is applied to help the model generalise better. The output of the model is a feature vector of size \num{128}.

\paragraph{BiLSTM-based.}
The model has an architecture similar to LogRobust, which was proposed for anomaly detection. Due to its RNN-based architecture, its output is a feature vector with the same size as the input log sequence length~\cite{LogRobust}. 

\paragraph{CNN-based.}\label{sec:arch:DL:CNN}
The CNN architecture is a variation of the convolutional design for the CNN-based anomaly detection mode~\cite{CNN}. Based on our preliminary experimental results, 20 filters, instead of one, for each of the three 1D convolutions (see \S~\ref{sec:background:cnn}) are used in parallel to capture relationships between log templates at different distances. Padding is used to ensure that feature maps of each convolution have the same dimension as the input. Hence, the length of the output feature vector is the product of the number of filters (20), the number of convolutions (3), and the input size of the log sequence.

\paragraph{Transformer-based.}
Our architecture of the transformer model is inspired by recent work in anomaly detection~\cite{Neurallog, HitAnomaly, LogSy}. The model is composed of two main parts: positional embedding and transformer blocks. One transformer block is adopted after positional embedding, set similarly to a recent study~\cite{Neurallog}. After global average pooling, the output matrix is mapped into one feature vector of the same size as the log template embedding $\theta = 768$, previously explained in \S~\ref{sec:background:ml-based-log-analysis}.

\section{Empirical Study Design}\label{sec:empirical-design}

\subsection{Research Questions}\label{sec:RQs}
The goal of this study is to systematically evaluate the performance of failure predictors, by instantiating our base architecture with different configuration of DL encoders and log sequence embedding strategies, for various datasets with different characteristics. The ultimate goal is to rely on such analyses \emph{to provide practical guidelines to select the right failure prediction model based on the characteristics of a given dataset}. 
To achieve this, we investigate the following research questions:
\begin{enumerate}[\bf RQ1:]

\item What is the impact of different DL encoders on failure prediction accuracy?
\item What is the impact of different log sequence embedding strategies on failure prediction accuracy?
\item How do DL-based failure predictors fare compared to traditional ML-based ones in terms of failure prediction accuracy?
\item What is the impact of different dataset characteristics on failure prediction accuracy? 

\item How does the accuracy of failure prediction on synthesised datasets compare to that of real-world datasets?
\end{enumerate}

RQ1 and RQ2 investigate how failure prediction accuracy varies across DL encoders and embedding strategies reported in the literature. Most of them have been evaluated in isolation or with respect to a few alternatives, often using ad-hoc benchmarks (see \S~\ref{sec:related-works} for a detailed comparison). 
To address this, we comprehensively consider all variations of our base architecture, obtained by combining different DL encoders and log sequence embedding strategies

that have been widely used in failure prediction and anomaly detection. 
Furthermore, we systematically vary the characteristics of the input datasets in terms of the number of log sequences, the length of log sequences, and the proportion of normal log sequences.  
The answer to these questions is expected to lead to practical guidelines for choosing the best failure prediction model given a dataset with certain characteristics.  

RQ3 compares the DL-based and traditional ML-based (also referred to as non-DL) failure predictors in terms of accuracy. This will allow us to better understand the potential advantages and drawbacks of using DL methods for failure prediction.

RQ4 additionally investigates the impact of the input dataset characteristics on failure prediction accuracy with a focus on the best DL encoder and log sequence embedding strategy found in RQ1 and RQ2.
The answer to this question will help us better understand under which conditions the configuration  of the best DL encoder and log sequence embedding strategy works sufficiently well for practical use, possibly leading to practical guidelines to best prepare input datasets for increasing failure prediction accuracy.

RQ5 compares the results (in terms of failure prediction accuracy) obtained by the configuration of the best DL encoder and log sequence embedding strategy on synthetic data with those obtained on a real dataset (more details in~\S~\ref{sec:real-world-dataset-processing}). 

\subsection{Methodology}\label{sec:methodology}

As discussed in \S~\ref{sec:architecture}, we can instantiate the base architecture for failure prediction with different DL encoders and log sequence embedding strategies. 

To answer RQ1 and RQ2, we train different configurations of the base architecture while systematically varying training datasets' characteristics (e.g., size and failure types). Then, we evaluate the relative performance of the configurations in terms of failure prediction accuracy, using test datasets having the same characteristics but not used during training. We elaborate on the different configurations, dataset characteristics, and failure predictor training and testing in the following sections.

To answer RQ3, we compare the results of the best configuration of the DL-based failure prediction architecture with a traditional ML-based failure predictor. We selected Random Forest (RF) as a traditional ML-based method since, according to the comprehensive study of \citet{Fernandez-Delgado2014}, it has shown the best performance overall compared to other traditional ML-based methods. Moreover, in the context of log-based failure prediction, the RF-based method has shown better results compared to other traditional ML-based methods, such as xgboost and SVM~\cite{prefix, rfvsxgboost} (see also \S~\ref{sec:related-works:FP}).
Therefore, using RF provides the best insights over using DL-based failure predictors. For RF, we set the number of estimators, which is the primary hyper-parameter, to 10, in line with its related study~\cite{Wu2023}. For embedding strategy, since the input of RF is an embedding vector rather than an embedding matrix used for our modular architecture, we selected TF-IDF (template-level), the best overall embedding strategy for RF according to a close study~\cite{Wu2023}.

To answer RQ4, we first identify all the top configurations since there might be certain datasets where configurations other than the best configuration inferred from RQ1-2 fare better. We then analyse the impact of each dataset characteristic (e.g., dataset size, percentage of failure) on these configurations. 
To further investigate the combination of these characteristics, we construct a decision tree based on the best configuration for each dataset to predict the conditions where each top configuration fares best.
 
Moreover, we build regression trees~\cite{breiman1984classification} to automatically infer conditions describing how the failure prediction accuracy of the best configurations varies according to the dataset characteristics. 

To answer RQ5, due to the limited availability of real-world datasets for failure prediction (see~\S~\ref{sec:related-works:FP}), we must choose from the datasets available for anomaly detection. Especially, datasets designed explicitly for failure detection (i.e., a sub-task of anomaly detection) are more compatible with our task, considering their transferability to failure prediction discussed in~\S~\ref{sec:background:FP}. 
OpenStack is a common dataset explicitly used for failure prediction~\cite{Bogatinovski2022FailureIF} and further labelled at the log message level that we refer to as OpenStack\_v2. These characteristics allowed us to further process it to make it suitable for failure prediction, leading to the creation of a new dataset called OpenStack\_PF, which we introduce in~\S~\ref{sec:real-world-dataset-processing}. We compare the failure prediction accuracy results obtained on the synthesized datasets most similar, in terms of dataset size, failure percentage, and MLSL~\footnote{More details are provided in~\S~\ref{sec:res:rq5}.} to OpenStack\_PF, with those obtained on the OpenStack\_FP dataset. For practical reasons, we only focus on the accuracy results of the best DL configuration as well as the best traditional ML model, i.e., RF.

\subsubsection{Log Sequence Embedding Strategies and DL Encoders}\label{sec:design:DL&embed}
As for different log sequence embedding strategies, we considered the best-fitting instances from three categories, which have shown to be accurate in the literature as discussed in \S~\ref{sec:arch:LM}. 
Among template ID-based strategies, we excluded the count vector since they are unable to capture sequential patterns in a log sequence (see \S~\ref{sec:background:embed}). 
TF-IDF methods (like count vectors) were incompatible with our architecture since their output embedding is a vector for each log sequence rather than a matrix. Conversely, Logkey2vec incorporates the order of log templates in the embedding procedure and yields the desired output structure.
Among semantic-based strategies, since Template2vec is trained on manually added, domain-specific synonyms and antonyms, its applicability is limited and we excluded it, as mentioned in \ref{sec:background:bert}. Among available pre-trained strategies, we included BERT, given its prevalent usage in log analysis studies and its demonstrated benefits~\cite{logBERT, Neurallog}.
As for the hybrid strategy, we included F+T (aggregation of FastText with TD-IDF), which is a common hybrid strategy in the existing literature.

As for different DL encoders in RQ1 and RQ2, we consider four encoders (LSTM, BiLSTM, CNN, and transformer) that have been previously used in related works; we describe their architecture details in \S~\ref{sec:arch:DL}. 
We configured the encoders based on the recommendations reported in the literature (see \S~\ref{sec:arch:DL} for further details).

\subsubsection{Datasets with Different Characteristics}\label{sec:DatasetCharacteristics}
As for the characteristics of datasets, we consider four factors that are expected to affect failure prediction performance: 
(1) dataset size (i.e., the number of logs in the dataset), 
(2) log sequence length (LSL) (i.e., the length of a log sequence in the dataset), 
(3) failure percentage (i.e., the percentage of log sequences with failure patterns in the dataset), and
(4) failure pattern type (i.e., types of failures).

The dataset size is important to investigate to assess the training efficiency of different DL models. To consider a wide range of dataset sizes while keeping the number of all combinations of the four factors tractable, we consider six levels that cover the range of real-world dataset sizes reported in a recent study~\cite{10.1145/3510003.3510155}: \num{200}, \num{500}, \num{1000}, \num{5000}, \num{10000}, and \num{50000}. 

The LSL could affect failure prediction since a failure pattern that spans a longer log might be more difficult to predict correctly. Similar to observed lengths in real-world log sequences across publicly available datasets~\cite{10.1145/3510003.3510155}, we vary the maximum\footnote{We set the maximum LSL for to simplify control.} LSL across five levels: \num{20}, \num{50}, \num{100}, \num{500}, and \num{1000}.

The failure percentage determines the balance of classes in a dataset, which may affect the performance of DL models~\cite{ImbalanceLearning}. The training dataset is perfectly balanced at \num{50}\%. However, the failure percentage can be much less than \num{50}\% in practice, as observed in real-world datasets~\cite{unbalance_failure}. Therefore, we vary the failure percentage across six levels: \num{5}\%, \num{10}\%, \num{20}\%, \num{30}\%, \num{40}\%, and \num{50}\%. 

Regarding failure patterns, we aim to consider patterns with potential differences in terms of learning effectiveness. However, failure patterns defined in previous studies are too simple; for example, \citet{Das2018} consider a specific, consecutive sequence of problematic log templates, called a ``failure chain''. But in practice, not all problematic log templates appear consecutively in a log. To address this, we use regular expressions to define failure patterns, allowing non-consecutive occurrences of problematic log templates. For example, a failure pattern ``$x(y|z)$'' indicates a pattern composed of two consecutive templates that starts with template $x$ and ends with either template $y$ or template $z$. 
In addition, we consider two types of failure patterns (in the form of regular expressions), \textit{Type-F} and \textit{Type-I}, depending on the cardinality of languages accepted by the regular expressions (\textit{finite} and \textit{infinite}, respectively). This is because, if the cardinality of the language is finite, DL models might memorise (almost) all the finite instances (i.e., sequences of log templates) instead of learning the failure pattern. For example, the language defined by the regular expression ``$x(y|z)$'' is finite since there are only two template sequences (i.e., $xy$ and $xz$) matching the expression ``$x(y|z)$. In this case, the two template sequences might appear in the training set, making it straightforward for DL models to simply memorise them. On the contrary, the language defined by the regular expression ``$x^*(y|z)$'' is infinite due to infinite template sequences that can match the sub-expression `$x^*$'; therefore simply memorising some of the infinitely many sequences matching  ``$x^*(y|z)$'' would not be enough to achieve high failure prediction accuracy.

To sum up, we consider \num{360} combinations (six dataset sizes, five maximum LSLs, six failure percentages, and two failure pattern types) in our evaluation. 
However, we could not use publicly available datasets for our experiments due to the following reasons.
First, although \citet{He2021} reported several datasets in their survey paper, they are mostly labelled based on the occurrence of error messages (e.g., log messages with the level of ERROR) instead of considering failure patterns (e.g., sequences of certain messages). 
Furthermore, there are no publicly available datasets covering all the combinations of the four factors defined above, making it impossible to thoroughly investigate their impact on failure prediction.
To address this issue, we present a novel approach for synthetic log data generation in \S~\ref{sec:syn-data-gen}. 

\subsubsection{Real-world Dataset Processing}\label{sec:real-world-dataset-processing}
The real-world log dataset used to address RQ5 is based on the OpenStack dataset, which is collected from a large-scale study on failures in OpenStack, as documented by~\citet{howbadbug}. It is known to be the most comprehensive publicly available dataset of logs including failure data generated from a cloud-based system~\cite{Bogatinovski2022FailureIF}, involving a wide variety of failures reported in the OpenStack bug repository\footnote{\url{https://bugs.launchpad.net/openstack/}}. Failures stem from different fault injection mechanisms (e.g., modifying the source code of OpenStack) and running a workload (task) with the injected fault. In the original OpenStack dataset, the granularity of the labels is at the level of the workload; labels are determined by checking assertions at the end of the workload runs. \citet{Bogatinovski2022FailureIF} further labelled the logs at the log message level using two human annotators labelling more than \num{200000} log messages to find those indicating the logged failure. We name this version of the dataset OpenStack\_v2. To make OpenStack\_v2 ready for failure prediction, we further processed it according to the discussion on dataset transferability, as mentioned in~\S~\ref{sec:background:FP}.

Specifically, we partition the logs according to their log identifier, which is the task ID in this context. As discussed in~\S~\ref{sec:background:AD}, partitioning logs using log identifiers leads to higher accuracy than using timestamp-based ones. If a task ID is marked as a failure, we retain only the log messages, ordered by timestamp, up before the occurrence of the first failure message. In this way, we eliminate the direct signs of a failure in a log, resulting in a log sequence that appears normal although it triggers a failure. Additionally, due to the limitation on maximum log sequence length, in case a log sequence exceeds the limit, we only keep the last 1000 log messages. We set this threshold since it is the maximum input sequence length in our modular architecture; moreover, we speculate the messages at the end of the sequence to be more related to the subsequent failure. We name the processed dataset OpenStack\_PF, as it is suitable for failure prediction. Table~\ref{table:openstack} provides a summary of the OpenStack\_FP statistics, where ``\# logs'' indicates the number of logs that form a log sequence, and ``avg,'' ``min,'' and ``max'' represent the average, minimum, and maximum lengths of the log sequences, respectively.

\begin{table}
    \centering
          \resizebox{\columnwidth}{!}{
\begin{threeparttable}[t]
 \caption{Overview of OpenStack\_PF dataset}\label{table:openstack}

\begin{tabular}{ccccccc}
\toprule
\textbf{\#Logs} & \textbf{\makecell{\#Failures\\Log Sequences}} & \textbf{\makecell{Failure\\ Percentage}} & \textbf{\makecell{\#Unique \\Log Templates}}  & \textbf{\makecell{\\avg}} & \textbf{\makecell{Log Sequence Length\\min}} & \textbf{\makecell{\\max}}\\
\midrule
 876 & 188 & ~21.46\% & 468 & 228 & 4 & 462\\ 
\bottomrule
\end{tabular}
\end{threeparttable}}

\end{table}

\subsubsection{Failure Predictor Training and Testing}\label{sec:training-testing}
We split each artificially generated dataset, as well as OpenStack\_PF, into two disjoint sets, a training set and a test set, with a ratio of \num{80}:\num{20}. Further, \num{20}\% of the training set is separated as a validation set, which is used for early stopping~\cite{earlystopping} during training to avoid over-fitting. 

For training failure predictors, to control the effect of highly imbalanced datasets,  oversampling~\cite{upton2008dictionary} is performed on the minority class (i.e., failure logs) to achieve a \num{50}:\num{50} ratio of normal to failure logs in the training dataset.
For all the training datasets, we use the Adam optimizer~\cite{adam} with a learning rate of 0.001 and the sparse categorical cross-entropy loss function~\cite{chen2022ai} considering the Boolean output (i.e., failure or not) of the models.
However, we use different batch sizes and numbers of epochs for datasets with different characteristics since they affect the convergence speed of the training error (particularly the dataset size, the maximum LSL, and the failure percentage). It would however be impractical to fine-tune the batch size and the number of epochs for \num{360} individual combinations. Therefore, based on our preliminary evaluation results, we use larger batch sizes with fewer epochs for larger datasets to keep the training time reasonable without significantly affecting training effectiveness. 
Specifically, we set the two hyperparameters as follows:
\begin{itemize}[-]
    \item \textit{Batch size}: By default, we set it to \num{10}, \num{15}, \num{20}, \num{30}, \num{150}, and \num{300} for dataset sizes of \num{200}, \num{500}, \num{1000}, \num{5000}, \num{10000}, and \num{50000}, respectively. If the failure percentage is less than or equal to \num{30} (meaning more oversampling will happen to balance between normal and failure logs, increasing the training data size), then we increase the batch size to \num{10}, \num{15}, \num{30}, \num{60}, \num{300}, and \num{600}, respectively, to reduce training time. Furthermore, regardless of the failure percentage, we set the batch size to \num{5} if the maximum LSL is greater than or equal to \num{500} to prevent memory issues during training.
    \item \textit{Number of epochs}: By default, we set it to \num{20}. If the maximum LSL is greater than or equal to 500, we reduce the number of epochs to \num{10}, \num{10}, \num{5}, and \num{5} for dataset sizes of \num{1000}, \num{5000}, \num{10000}, and \num{50000}, respectively, to reduce training time.
\end{itemize}

\begin{table}
    \centering
         \begin{threeparttable}[t]
 \caption{Overview of Hyperparameter Setting}\label{table:batch-epoch}
\begin{tabular}{cccccccc}
\toprule
\multirow{2}{*}{\textbf{Hyperparameter}} & \multirow{2}{*}{\textbf{Condition}} & \multicolumn{6}{c}{\textbf{Dataset Size}}\\
& & \textbf{\num{200}} & \textbf{\num{500}} & \textbf{\num{1000}} & \textbf{\num{5000}} & \textbf{\num{10000}} & \textbf{\num{50000}}\\ 
\midrule
\multirow{3}{*}{Batch Size} & Default & 10 & 15 & 20 & 30 & 150 & 300\\
& $\text{PF} \leq 30$ & 10 & 15 & 30 & 60 & 300 & 600\\
& $\text{MLSL}\geq 500$\tnote{*} & 5 & 5 & 5 & 5 & 5 & 5\\ 
\midrule
\multirow{2}{*}{Number of Epochs} & Default & 20 & 20 & 20 & 20 & 20 & 20\\
& MLSL $\geq$ 500 & 20 & 20 & 10 & 10 & 5 & 5\\ 
\bottomrule
\end{tabular}
\begin{tablenotes}
         \item [*] This condition has higher priority than the other.
         \end{tablenotes}
\end{threeparttable}

\end{table}
Table~\ref{table:batch-epoch} summarises the above conditions, where FP is the failure percentage and MLSL refers to the maximum LSL. 
For OpenStack\_FP, we determined the hyperparameter settings by matching its characteristics to the closest ones in the table (i.e., dataset size of \num{1000}, failure percentage of 20\%, and MLSL of \num{500}).

Once failure predictors are trained, we measure their accuracy on the corresponding test set in terms of precision, recall, and F1 score. We also refer to robustness as a degree of consistency in accuracy in the presence of varying data set characteristics.

We conducted all experiments with cloud computing environments provided by the Digital Research Alliance of Canada~\cite{computecanada}, on the Cedar cluster with a total of \num{94528} CPU cores for computation and \num{1352} GPU devices.

\subsection{Synthetic Data Generation}\label{sec:syn-data-gen}

In defining a set of factors, the methodology described in \S~\ref{sec:methodology} makes it clear that there is a need for a mechanism that can generate datasets in a controlled, unbiased manner. Recent works on data synthesis have used finite-state automata (\S~\ref{sec:rw:data-syn}), but they cannot accommodate the set of factors that we aim to control during synthetic data generation.
For example, let us consider the factor of failure percentage (\S~\ref{sec:DatasetCharacteristics}). Such a factor requires that one be able to control whether the log sequence being generated does indeed correspond to a failure; this would ultimately allow one to control the percentage of failure log sequences in a generated dataset.

While, for smaller datasets, one could imagine manually choosing log sequences that represent both failures and normal behaviour, for larger datasets this is not feasible. When considering the other factors defined in \S~\ref{sec:methodology}, such as \emph{LSL}, the case for a mechanism for automated, controlled generation of datasets becomes yet stronger.

\subsubsection{Key Requirements}\label{sec:data-generator-reqs}

We now describe a set of requirements that must be met by whatever approach we opt to take for generating datasets. In particular, our approach should:

\newcounter{req}
\newcommand{\requirement}[2]{\refstepcounter{req}\paragraph{\label{#1}\textbf{R\thereq} - #2}}

\requirement{req:controlled}{Allow datasets' characteristics to be controlled.}  This requirement has already been described, but we summarise it here for completeness. We must be able to generate datasets for each combination of levels (of the factors defined in \S~\ref{sec:methodology}). Hence, our approach must allow us to choose a combination of levels, and generate a dataset accordingly.

\requirement{req:realistic}{Be able to generate realistic datasets.} A goal of this work is to present results that are applicable to real-world systems. Hence, we must require that the datasets with which we perform any evaluations reflect real-world system behaviours.

\requirement{req:diverse}{Be able to generate datasets corresponding to a diverse set of systems.}  While we require that the datasets that we use be realistic, we must also ensure that the data generator can generate log sequences for any system, rather than being limited to a single system.

\requirement{req:unbiased}{Avoid bias in the log sequences that make up the generated datasets.}

For a given system, we wish to generate datasets containing log sequences that explore as much of the system's behaviour as possible (rather than being biased to a particular part of the system).

\subsubsection{Automata for System Behaviour}\label{sec:automata}

Our approach is based on finite-state automata. In particular, we use automata as approximate models of the behaviour of real-world systems. We refer to such automata as \emph{behaviour models}, since they represent the computation performed by (i.e., behaviour of) some real-world system. We chose automata, or behaviour models, because some of our requirements are met immediately:

\paragraph{\textbf{R\ref{req:realistic}}.}  Existing tools~\cite{PRINS, MINT} allow one to infer behaviour models of real-world systems from collections of these systems' logs (in a process called \emph{model inference}). Such models attach log messages to transitions, which is precisely what we need. Importantly, collections of logs used are unlabelled, meaning that the models that we get from these tools have no existing notion of normal behaviour or failures.

\paragraph{\textbf{R\ref{req:diverse}}.}  A result of meeting R\ref{req:realistic} is that one can easily infer behaviour models for multiple systems, provided the logs of those systems are accessible.

\paragraph{\textbf{R\ref{req:unbiased}}.}  If we are to use automata to represent systems, then we can define bias of collections of log sequences in terms of \emph{how much} of a behaviour model is represented in those log sequences.

\mbox{}\\
The remaining sections will give the complete details of our automata-based data generation approach. In presenting these details, we will show how R\ref{req:controlled} and R\ref{req:unbiased} are met.

\subsubsection{Behaviour Models}\label{sec:behavioural-models}
We take a behaviour model $\mathcal{M}$ to be a deterministic finite-state automaton $\langle Q, A, q_0, \Sigma, \delta \rangle$, with symbols as defined in \S~\ref{sec:background:automata}.

A behaviour model has the particular characteristic that its alphabet $\Sigma$ consists of \emph{log template IDs} (see \S~\ref{sec:background:logs}). A direct consequence of this is that one can extract log sequences from behaviour models. In particular, if one considers a sequence of states (i.e., a path) $q_0, q_i, q_{i+1}, \dots, q_n$ through the model, one can extract a sequence of log template IDs using the transition function $\delta$. For example, if the first two states of the sequence are $q_0$ and $q_i$, then one need only find $s \in \Sigma$ such that $\delta(q_0, s) = q_i$, i.e., it is possible to transition to $q_i$ from $q_0$ by observing $s$.
Finally, by replacing each log template ID in the resulting sequence with its corresponding log template, one obtains a \emph{log sequence} (see \S~\ref{sec:background:logs}). These sequences can be divided into two categories: \emph{failure log sequences} and \emph{normal} log sequences. 

We describe failures using regular expressions. This is natural since behaviour models are finite state automata, and sets of paths through such automata can be described by regular expressions.
Hence, we refer to such a regular expression as a \emph{failure pattern}, and denote it by $\mathit{fp}$. By extension, for a given behaviour model $\mathcal{M}$, we then denote by $\mathsf{failurePatterns}(\mathcal{M})$ the set $\{\mathit{fp}_1, \mathit{fp}_2, \dots, \mathit{fp_n}\}$ of failure patterns paired with the model $\mathcal{M}$. Based on this, we characterise \emph{failure log sequences} as such:

\paragraph{Failure log sequence.} For a system whose behaviour is represented by a behaviour model $\mathcal{M}$, we say that a log sequence represents a failure of the system whenever its sequence of log template IDs matches some failure pattern $\mathit{fp} \in \mathsf{failurePatterns}(\mathcal{M})$.

\mbox{}\\
Since this definition of failure log sequences essentially captures a subset of the possible paths through $\mathcal{M}$, we define normal log sequences as those log sequences that are not failures:

\paragraph{Normal log sequence.} For a system whose behaviour is represented by a behaviour model $\mathcal{M}$, we say that a log sequence $l$ is normal, i.e., it represents normal behaviour, whenever $l \in \mathcal{L}(\mathcal{M})$ and $l \not\in \bigcup_{\mathit{fp} \in \mathsf{failurePatterns}(\mathcal{M})}{\mathcal{L}(\mathit{fp})}$ (we take $\mathcal{L}(\mathcal{M})$ and $\mathcal{L}(\mathit{fp})$ to be as defined in \S~\ref{sec:background:automata}). Hence, defining a normal log sequence requires that we refer to both the language of the model $\mathcal{M}$, and the languages of all failure patterns associated with the model $\mathcal{M}$.

\subsubsection{Generating Log Sequences for Failures}\label{sec:generating-failures}

Let us suppose that we have inferred a model $\mathcal{M}$ from the execution logs of some real-world system, and that we have defined the set $\mathsf{failurePatterns}(\mathcal{M})$. Then we generate a failure log sequence that matches some $\mathit{fp} \in \mathsf{failurePatterns}(\mathcal{M})$ by:
\begin{enumerate}
    
    \item Computing a subset of $\mathcal{L}(\mathit{fp})$. 
    We do this by repeatedly generating single members of $\mathcal{L}(\mathit{fp})$. Ultimately, this leads to the construction of a subset of $\mathcal{L}(\mathit{fp})$. In practice, the Python package \textit{exrex}\cite{exrex} can be used to generate random words from the language $\mathcal{L}(\mathit{fp})$, so we invoke this library repeatedly.
    
    If the language of the regular expression is infinite, we can run \textit{exrex} multiple times, each time generating a random string from the language. The number of runs is set based on our preliminary results with respect to the range of dataset size (2500 times for each failure pattern). Doing this, we generate a subset of $\mathcal{L}(\mathit{fp})$.

    \item Choosing at random a log sequence $l$ from the random subset $\mathcal{L}(\mathit{fp})$ computed in the previous step, with $|l| \le \mathit{mlsl}$ where $mlsl$ refers to the value of maximum log sequence manipulated by LSL factor. (see \S~\ref{sec:background:logs}) (\emph{maximum LSL}, see \S~\ref{sec:methodology}). The Python package \textit{random}\cite{random} was employed for this.

    We highlight that failure patterns are designed so that there is always at least one failure pattern that can generate log sequences whose length falls within this bound.

    More details on this are provided in \S~\ref{sec:expr-setting}.
\end{enumerate}

For requirement R\ref{req:unbiased}, since our approach relies on random selection of log sequences from languages generated by the \emph{exrex} tool, we highlight that the bias in our approach is subject to the implementation of both \emph{exrex}, and the Python package, \textit{random}. \emph{Exrex} is a popular package for RegEx that has more than 100k monthly downloads. Its method for generating a random matching sequence is implemented by a random selection of choices on the RegEx's parse tree nodes.
\textit{Random} package uses the Mersenne twister algorithm~\cite{Matsumoto1998} to generate a uniform pseudo-random number used for random selection tasks. 

\subsubsection{Generating Log Sequences for Normal Behaviour}\label{sec:generating-nominal}

While our approach defines how failures should look using a set of failure patterns $\mathsf{failurePatterns}(\mathcal{M})$ defined over a model $\mathcal{M}$, we have no such definition of how normal behaviour should look. Instead, this is left implicit in our behaviour model. However, based on the definition of normal log sequences given in \S~\ref{sec:behavioural-models}, such log sequences can be randomly generated by performing random walks on behaviour models.

This fact forms the basis of our approach to generating log sequences for normal behaviour. However, we must also address key issues:  1) the log sequences generated by our random walk must be of bounded length, and 2) the log sequences must also lack bias.

There are two reasons for enforcing a bound on the length of log sequences:
\begin{itemize}
    \item Deep learning models (such as CNN) often accept inputs of limited size, so we have to ensure that the data we generate is compatible with the models we use.
    \item One of the factors introduced in \S~\ref{sec:methodology} is LSL, so we need to be able to control the length of log sequences that we generate.
\end{itemize}

For bias, we have two sources: 1) bias to specific regions of the behavioural model, 2) bias to limited variation of LSL. We must minimise bias in both cases.

Algorithm~\ref{alg:generate-nominal-sequence} gives our procedure for randomly generating a log sequence representing normal behaviour of a system. Algorithm~\ref{alg:generate-nominal-sequence} itself makes use of Algorithm~\ref{alg:random-walk}.

\begin{algorithm}[!ht]
\DontPrintSemicolon
  \SetKwInput{KwInput}{Input}                
\SetKwInput{KwOutput}{Output}  
  \KwInput{$ \mathcal{M}: \mathit{behaviour\ model}, \mathit{mlsl} : int$}
  \KwOutput{$ sequence: \langle s_1, s_2, \dots, s_n\rangle \in \mathcal{L}(\mathcal{M})$}
  $\mathit{sequence: list} \leftarrow \mathit{filteredRandomWalk}(\mathcal{M}, \mathit{mlsl}) $
  
  \While{$\mathit{sequence} \in \bigcup_{\mathit{fp}_i \in \mathsf{failurePatterns}(\mathcal{M})}\mathcal{L}(\mathit{fp}_i)$}
   {
   		$\mathit{sequence} \leftarrow \mathit{filteredRandomWalk}(\mathcal{M}, \mathit{mlsl})$\;
   }
  \textbf{return} $\mathit{sequence}$

\caption{$\mathit{generateNormalSequence}$\label{alg:generate-nominal-sequence}}
\end{algorithm}

\begin{algorithm}
    \DontPrintSemicolon
  \SetKwInput{KwInput}{Input}                
\SetKwInput{KwOutput}{Output}  
  \KwInput{$\mathcal{M} = \langle Q, A, q_0, \Sigma, \delta \rangle : \mathit{behaviour\ model}, \mathit{mlsl}: int$}
  \KwOutput{$sequence : \langle s_1, s_2, \dots, s_n\rangle\in \mathcal{L}(\mathcal{M}) $}
  $\mathit{sValue}: int \leftarrow \mathit{calculateSValues}(\mathcal{M})$\\
   $\mathit{sequence}: list$ $\leftarrow$ $\langle \rangle$\\
  $\mathit{maximumWalk} : int\leftarrow \mathit{mlsl}$\\
  $\mathit{currentState: state}$ $\leftarrow$ $q_0$\\
  
  \While{$\mathit{currentState} \notin A$}
   {
   		$\mathit{options} : set \leftarrow \emptyset$\\
     $\mathit{transitions: set} \leftarrow \{\langle \mathit{currentState}, s, q \rangle : \delta(\mathit{currentState}, s) = q\}$\;
     \For{$\langle q, s, q' \rangle$ $\in$ transitions}
     {
     \If{$\mathit{sValue}(q') < \mathit{maximumWalk}$}
     {
     $\mathit{options} \leftarrow \mathit{options} \cup \{\langle q, s, q' \rangle \}$
     }
     }
     $\langle q, s, q' \rangle\leftarrow \textit{random choice from } \mathit{options} $\\
    $\mathit{sequence} \leftarrow \mathit{sequence} + \langle s \rangle$\\
    $\mathit{currentState} \leftarrow q'$\\
    $\mathit{maximumWalk} \leftarrow \mathit{maximumWalk} - 1$
   }
   \textbf{return} $\mathit{sequence}$

\caption{$\mathit{filteredRandomWalk}$\label{alg:random-walk}}
\end{algorithm}

In particular, Algorithm~\ref{alg:generate-nominal-sequence} generates a normal log sequence by:
\begin{enumerate}
\item Generating a random log sequence by random walk (invoking Algorithm~\ref{alg:random-walk});
\item Looking for a failure pattern $\mathit{fp} \in \mathsf{failurePatterns}(\mathcal{M})$ that matches the generated log sequence;
\item Repeating until a log sequence is generated that matches no failure pattern.
\end{enumerate}
Ultimately, Algorithm~\ref{alg:generate-nominal-sequence} is relatively lightweight; the weight of the work is performed by Algorithm~\ref{alg:random-walk}, which we now describe in detail.

The input arguments of Algorithm~\ref{alg:random-walk}, which defines the procedure \textit{filteredRandomWalk}, are a behaviour model $\mathcal{M}$ and the maximum LSL, $\mathit{mlsl}$.

The algorithm proceeds as follows. First, on line 1, we invoke the \textit{calculateSValues} function to compute a map that sends each state $s \in Q$ of $\mathcal{M}$ to the length of the shortest path from that state to an accepting state in $A$. Next, on line 2, the \textit{sequence} variable is initialised to an empty sequence. As the algorithm progresses, this variable stores the generated sequence of log template IDs. To help with this, the variable \textit{currentState} is initialised to keep track of the state that the algorithm is \emph{currently in} during the walk of the behavioural model. Hence, this variable is initialised on line 4 as the initial state. The final step in the setup stage of our algorithm is to initialise the \textit{maximumWalk} variable, which serves as a counter to ensure the limit on the length of the generated log sequence (defined by $\mathit{mlsl}$) is respected. 

In the while loop (line 5), as long as the current state, $\mathit{currentState}$, is not yet an accepting state, the random walk transits from the current state to a new state. The set of possible transitions to take is computed by line 7, and stored in the variable $\mathit{transitions}$. Each transition is represented by a triple containing the starting state, the symbol to be observed, and the state resulting from observation of that symbol. Once this set has been computed, the algorithm performs a filtering step. In particular, in order to ensure that we respect the limit imposed on the length of the generated path by $\mathit{mlsl}$, we only consider transitions that lead to a state $q'$ such that $\mathit{sValue}(q') < \mathit{maximumWalk}$. The resulting list of valid options is then held in the variable $\mathit{options}$.

Once the set $\mathit{options}$ has been computed, one transition $\langle q, s, q' \rangle$ will be chosen randomly from the set (line 13). This random choice eliminates bias because, each time we choose the next state to transition to, we do not favour any particular state (there is no weighting involved). This, extended over an entire path, means that we do not favour any particular region of a behaviour model. Now, from the randomly chosen transition $\langle q, s, q' \rangle$, the log template ID, $s$, is added to \textit{sequence} (via sequence concatenation); \textit{currentState} is updated to the next state, $q'$; and \textit{maximumWalk} is decreased by one. Based on the condition of the while loop (line 5), when $\mathit{currentState} \in A$ (i.e., the algorithm has reached an accepting state), the generated sequence $\mathit{sequence}$ is returned.

While Algorithm~\ref{alg:random-walk} generates an unlabelled log sequence, Algorithm~\ref{alg:generate-nominal-sequence} generates a normal log sequence. To do this, it starts by generating a log sequence, by invoking the \textit{filteredRandomWalk} procedure (Algorithm~\ref{alg:random-walk}). Since the sequence generated by Algorithm~\ref{alg:random-walk} is unlabelled, we must ensure that we do not generate a failure log sequence. We do this by checking whether the generated log sequence, $\mathit{sequence}$, belongs to the language of any failure pattern in $\mathsf{failurePatterns}(\mathcal{M})$. If this is indeed the case, another sequence must be generated. This process is repeated (line 2) until the log sequence generated by the call of $\mathit{filteredRandomWalk}$ does not match any failure pattern in $\mathsf{failurePatterns}(\mathcal{M})$. Once a failure log sequence has been generated, it is returned.

We acknowledge that this process could be inefficient (since we are repeatedly generating log sequences until we get one with the characteristics that we need). However, we highlight that failure patterns describe only a small part of a behaviour model (this is essentially the assumption that failure is a relatively uncommon event in a real system). Hence, normal log sequences generated by random walks can be generated without too many repetitions.

\subsubsection{Correctness and Lack of Bias}\label{sec:syn-correctness} We now provide a sketch proof of the correctness of Algorithm~\ref{alg:random-walk}, along with an argument that the algorithm eliminates bias.

To prove correctness, we show that, for a behaviour model $\mathcal{M}$, the algorithm always generates a sequence of log template IDs that correspond to the transitions along a path through $\mathcal{M}$.

The algorithm begins at $q_0$, by setting $\mathit{currentState}$ to $q_0$ (line 4). From $q_0$, and each successive state in the path, the possible next states must be adjacent to $\mathit{currentState}$ (line 7). Hence, the final value of $\mathit{sequence}$ after the while-loop at line 5 must be a sequence of log template IDs that correspond to the transitions along a path through $\mathcal{M}$.

Further, we must show that the sequence of log template IDs generated does not only correspond to a path through the behaviour model, but is of length at most $\mathit{mlsl}$ (one of the inputs of Algorithms~\ref{alg:random-walk} and~\ref{alg:generate-nominal-sequence}). This is ensured by three factors:
\begin{itemize}
\item The initialisation of the variable $\mathit{maximumWalk}$ on line 3.
\item The subsequent decrease by one of that variable each time a new log template ID is added to $\mathit{sequence}$.
\item Filtering of the possible next states in the random walk on line 9. In particular, on line 9 we ensure that, no matter which state we transition to, there will be a path that 1) leads to an accepting state; and 2) has length less than $\mathit{maximumWalk}$.
\end{itemize}
Finally, bias is minimised by two factors:
\begin{itemize}
    \item On line 13, we choose a random next state. Of course, here we rely on the implementation of random choice that we use.
    \item On line 9, while we respect the maximum length of the sequence of log template IDs, we do not enforce that we reach this maximum. Hence, we can generate paths of various lengths.
\end{itemize}

\begin{figure}[t]
\centering
\begin{tikzpicture}[shorten >=1pt,node distance=2cm,on grid,auto] 
   \node[state,initial] (q_0)   {$q_0$}; 
   \node[state] (q_1) [above right=of q_0] {$q_1$}; 
   \node[state] (q_2) [below right=of q_0] {$q_2$}; 
   \node[state,accepting](q_3) [below right=of q_1] {$q_3$};
    \path[->]                                               
    (q_0) edge  [bend left, above] node  {c,d} (q_1)
          edge  [bend left, above] node [swap] {a,b} (q_2)
    (q_1) edge  node  {c,d} (q_3)
          edge [loop above] node {b} ()
          edge  [bend left, above] node   {a} (q_0)
    (q_2) edge  node [swap] {a} (q_3) 
          edge [loop below] node {c} ()
          edge [bend left, above] node {b} (q_0)
          edge [bend right,left] node {d} (q_1);
\end{tikzpicture}
\caption{An example of a behaviour model.\label{fig:behaviour-model-example}}
\end{figure}
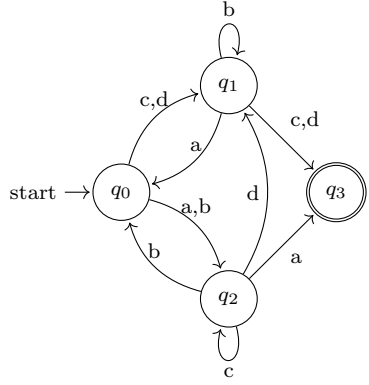

\paragraph{Example.} To demonstrate Algorithm~\ref{alg:random-walk}, we now perform a random walk over the behaviour model shown in Figure~\ref{fig:behaviour-model-example}. We start with the behaviour model's starting state, $q_0$, with $\mathit{mlsl}$ set to 5. Since $q_0 \not\in A$, we can execute the body of the while-loop at line 5. Hence, we can determine the set $\mathit{transitions}$ of transitions leading out of $q_0$:
\begin{equation*}
\{ \langle q_0,a,q_2\rangle, \langle q_0,b,q_2\rangle, \langle q_0,c, q_1\rangle, \langle q_0,d,q_1\rangle\}.
\end{equation*}
Our next step is to filter these transitions to ensure that the state that we move to allows us to reach an accepting state within $\mathit{maximumWalk}$ states. To do this, we filter the set $\mathit{transitions}$ with respect to the values in Table~\ref{svalue}. After this filtering step, the resulting set, $\mathit{options}$, is
\begin{equation*}
\{ \langle q_0,a,q_2\rangle, \langle q_0,b,q_2\rangle, \langle q_0,c, q_1\rangle, \langle q_0,d,q_1\rangle\}.
\end{equation*}
All states in $\mathit{transitions}$ are safe to transition to. To take one transition as an example, $\langle q_0, a, q_2\rangle$ has $\mathit{sValue}(q_2) = 1 < 5$, so is kept.

Once we have computed $\mathit{options}$, we choose a transition at random. In this case, we arrive at $\langle q_0,c,q_1\rangle$, meaning that we set $\mathit{currentState}$ to $q_1$. Before we progress to the next iteration of the main loop of the algorithm, we also decrease $\mathit{maximumWalk}$. This means that, during the next iteration of the while loop, we will be able to choose transitions leading to states from which an accepting state must be reachable within less than 4 states.

Indeed, from $q_1$, there are four transitions, for which we compute the set
\begin{equation*}
\{ \langle q_1,a,q_0\rangle, \langle q_1,b,q_1\rangle, \langle q_1,c, q_3\rangle, \langle q_1,d,q_3\rangle\}.
\end{equation*}
From this set, each possible next state has $\mathit{sValue}$ greater than $\mathit{maximumWalk}$ (equal to 4), so all of them would be possible options for the next step. Suppose that we choose $\langle q_1,a,q_0\rangle$ at random. Hence, $q_0$ is the next state and $a$ is added to the $\mathit{sequence}$. For the remaining steps, a possible run of the procedure could yield the sequence of transitions $\langle q_0,b,q_2\rangle$, $\langle q_2,d,q_1\rangle$, $\langle q_1,d,q_3\rangle$, in which case the final sequence of log template IDs would be $c, a, b, d, d$.

\begin{table}[t]
    \centering
    \caption{s values for each state}
    \begin{tabular}{|c|c|c|}
        \hline
         \textbf{state}& \textbf{s value}\\
         \hline
         \hline
         $q_0$& 2\\
         $q_1$  & 1 \\
         $q_2$  & 1\\
         $q_3$& 0\\
        \hline
    \end{tabular}
    \label{svalue}
\end{table}

\subsubsection{Compliance to Requirements}

We now describe how the approach that we have described meets the remaining requirements set out in \S~\ref{sec:data-generator-reqs}.

R\ref{req:controlled} is met because we have two procedures for generating failure log sequences (\S~\ref{sec:generating-failures}) and normal log sequences (\S~\ref{sec:generating-nominal}). By having these procedures, we can precisely control the number of each type of log sequence in our dataset.

R\ref{req:unbiased} is met because of the randomisation used in our data generation algorithm, described in Sections~\ref{sec:generating-failures} and~\ref{sec:generating-nominal}.


\subsection{Experimental Setting for Synthesised Data Generation}\label{sec:expr-setting}

To generate diverse log datasets with the characteristics described in \S~\ref{sec:DatasetCharacteristics}, using the syntactic data generation approach described in \S~\ref{sec:syn-data-gen}, we need two main artifacts: \textit{behaviour models} and \textit{failure patterns}.

\subsubsection{Behaviour Models}\label{sec:b-models}

Regarding behaviour models, as discussed in \S~\ref{sec:automata}, we can infer accurate models of real-world systems from their execution logs using state-of-the-art model inference tools, i.e., MINT~\cite{MINT} and PRINS~\cite{PRINS}. 
Among the potential models we could generate using the replication package of these tools, we choose models that satisfy the following criteria based on the model size and inference time reported by \citet{PRINS}:
\begin{enumerate}[(1)]
    \item The model should be able to generate (accept) a log with a maximum length of \num{20} (i.e., the shortest maximum LSL defined in \S~\ref{sec:DatasetCharacteristics};
    \item Since there is no straightforward way of automatically generating failure patterns for individual behaviour models considering the two failure pattern types, we had to manually generate failure patterns (detailed in \S~\ref{sec:fps}). Therefore, the size of the model should be amenable to manually generating failure patterns by taking into account the model structure (i.e., the number of all states and transitions is less than \num{1000});
    \item The model inference time should be less than 1 hour; and
    \item If we can use both PRINS and MINT to infer a model that satisfies the above criteria for the same logs, then we use PRINS, which is much faster than MINT in general, to infer the model.
\end{enumerate}
As a result, we use the following three models as our behaviour models: 
$\mathcal{M}_1$ (generated from NGLClient logs using PRINS), 
$\mathcal{M}_2$ (generated from HDFS logs using MINT), and 
$\mathcal{M}_3$ (generated from Linux logs using MINT). 
Table~\ref{table:models} reports about the size of the three behaviour models in terms of the number of templates (\#Templates), average length of templates (Avg Template Length) using a white-space separator, the number of states (\#States), and transitions (\#Transitions). 
It additionally shows the number of states in the largest strongly connected component (\#States-NSCC)~\cite{SCC}, which indicates the complexity of a behaviour model (the higher, the more complex).

\begin{table}
    \centering
    \caption{Overview of Behavioural models}\label{table:models}
      \resizebox{\columnwidth}{!}{
    \begin{threeparttable}
    \begin{tabular}{crrrrr}
        \toprule
        \textbf{Model}& \textbf{\#Templates} & \textbf{Avg Template Length} & \textbf{\#States} & \textbf{\#Transitions} & \textbf{\#States-NSCC}\\
        \midrule
        $\mathcal{M}_1$ & 70 & 54 & 154 & 195 & 5 \\
        $\mathcal{M}_2$ & 16 & 51 & 91 & 189 & 72 \\
        $\mathcal{M}_3$ & 115 & 39 & 350 & 486 & 331\\
        \bottomrule
    \end{tabular}
    \end{threeparttable}}
\end{table}

\subsubsection{Failure Patterns}\label{sec:fps}
Regarding failure patterns, recall a failure pattern $\mathit{fp}$ of a behaviour model $\mathcal{M}$ is a regular expression where $\mathcal{L}(\mathit{fp}) \subset \mathcal{L}(\mathcal{M})$, as described in \S~\ref{sec:behavioural-models}. 
Also, note that we need two types of failure patterns (\textit{Type-F} and \textit{Type-I}), and the failure log sequences generated from the failure patterns must satisfy the dataset characteristics (especially the maximum LSL) defined in \S~\ref{sec:DatasetCharacteristics}. 
To manually create such failure patterns (regular expressions) in an unbiased way, we used the following steps for each behaviour model and failure pattern type:
\begin{enumerate}[Step 1:]
    \item We randomly choose the alphabet size of a regular expression and the number of operators (i.e., alternations and Kleene stars; the latter is not used for \textit{Type-F}).
    \item Using the chosen random values, for a given behaviour model $\mathcal{M}$, we manually create a failure pattern (regular expression) $\mathit{fp}$ to satisfy $\mathcal{L}(\mathit{fp}) \subset \mathcal{L}(\mathcal{M})$ and the maximum LSL within the time limit of \num{1} hour; if we fail (e.g., if the shortest log in $\mathcal{L}(\mathit{fp})$ is longer than the maximum LSL of \num{20}), we go back and restart Step \num{1}.
    \item We repeat Steps \num{1} and \num{2} ten times to generate ten failure patterns and then randomly select three out of them.
\end{enumerate}
As a result, we use \num{18} failure patterns (i.e., \num{3} failure patterns $\times$ \num{3} behaviour models $\times$ \num{2} failure pattern types) for synthetic data generation.
Table~\ref{table:failurepatterns} reports the characteristics of failure patterns in terms of their behavioural model (Model), pattern type (Type), the length of the pattern in terms of letters and operators (Length), size of the alphabet (\#Alphabet), and the number of operators (\#Operators). Additionally, it includes the maximum depth of Kleene Star Structure(s) for \textit{Type-I} (Star Depth), which indicates the maximum depth of a nesting structure (e.g., 3 for $((b^*c)^*a)^*b$). Since there are three failure patterns per behavioural model and type, their values are presented in the form of a triple, respectively. For instance, under the $\mathcal{M}_2$ model and \textit{Type-I}, the first failure pattern has a length of 31 and an alphabet size of 5, uses 10 operators, and showcases a star depth of 1. While the complexity of failure patterns is bounded by their behavioural model (see~\S~\ref{sec:generating-failures}), there is a wide variability among failure patterns across each characteristic.

\begin{table}[t]
    \centering
    \caption{Overview of Failure Patterns}\label{table:failurepatterns}
    \begin{threeparttable}[t]
    \begin{tabular}{ccrrrr}
        \toprule
        \textbf{Model}& \textbf{Type} & \textbf{Length} & \textbf{\#Alphabet} & \textbf{\#Operators} & \textbf{Star Depth}\\
        \midrule
        \multirow{2}{*}{$\mathcal{M}_1$} & F & (14, 34, 35) & (7, 17, 30) & (5, 8, 1) & -\\
         & I &  (17, 25, 41) & (16, 15, 16) & (1, 7, 8) & (1, 2, 2) \\
         \midrule
        \multirow{2}{*}{$\mathcal{M}_2$} & F & (20, 27, 32) & (11, 9, 11) & (3, 6, 7) & - \\
         & I & (31, 8, 39) & (5, 5, 12) & (10, 1, 9) & (1, 1, 2)\\
         \midrule
        \multirow{2}{*}{$\mathcal{M}_3$} & F &  (134, 36, 48) & (77, 16, 14) & (16, 10, 5) & -\\
         & I &  (44, 30, 124) & (11, 16, 78) & (12, 7, 13) & (1, 2, 1) \\
        \bottomrule
    \end{tabular}
    \end{threeparttable}
\end{table}

\subsubsection{A Remark on Generalisability.}\label{sec:Generalisability} At this point, we highlight that we cannot give failure patterns that are representative of real-world patterns for two key reasons:
\begin{itemize}
\item Failure patterns are necessarily dependent on the behaviour model, itself representing a specific system.
\item To the best of our knowledge, there are no failure patterns derived from real-world systems reported in the literature.
\end{itemize}
Hence, instead of aiming to generate a set of failure patterns that are somehow representative of a target that is necessarily elusive, we aim to work with failure patterns that are \emph{diverse}.

We ensure this by first separating failure patterns into two types: \emph{Type-F} and \emph{Type-I}. Distinguishing between failure patterns that accept infinite and finite languages allows us to see how our failure prediction machinery performs when the language of log sequences to work with is infinite vs. finite.

Second, in varying the alphabet size, we control how much of a behaviour model a failure pattern can capture. Hence, across ten randomly generated failure patterns, we would generate failure patterns that could explore only a small region of the behaviour model, along with others that would explore larger regions of the behaviour model.

Further, in varying how many (if any) alternations are used, we control how many \emph{selections} can be made when traversing a behaviour model. For example, the failure pattern $a \mid b$ allows a single selection; we either take the $a$ transition, or the $b$ transition. However, the failure pattern $(a \mid b)(c \mid d)$ allows two selections; we first either take $a$ or $b$, and then we either take $c$ or $d$.

Finally, in varying how many (if any) Kleene stars are used, we control how many opportunities for \emph{cycles} we have when traversing our behaviour model. For example, if we have $a^*$, then we have a single opportunity to loop on $a$. If we have $(a \mid b)^*(c \mid d)^*$, then we have two opportunities to loop: on either $a$ or $b$, and then on either $c$ or $d$.

As a result, the various elements of control that we introduced above 

lead to the selection of failure patterns that will generate a large variety of log sequences from a single behaviour model.

\subsubsection{Overview of Synthesised Data.}\label{sec:syn-statis} 
\begin{figure}
\centering
  
  \begin{subfigure}{.45\linewidth}
    \centering
    \includegraphics[width = \linewidth]{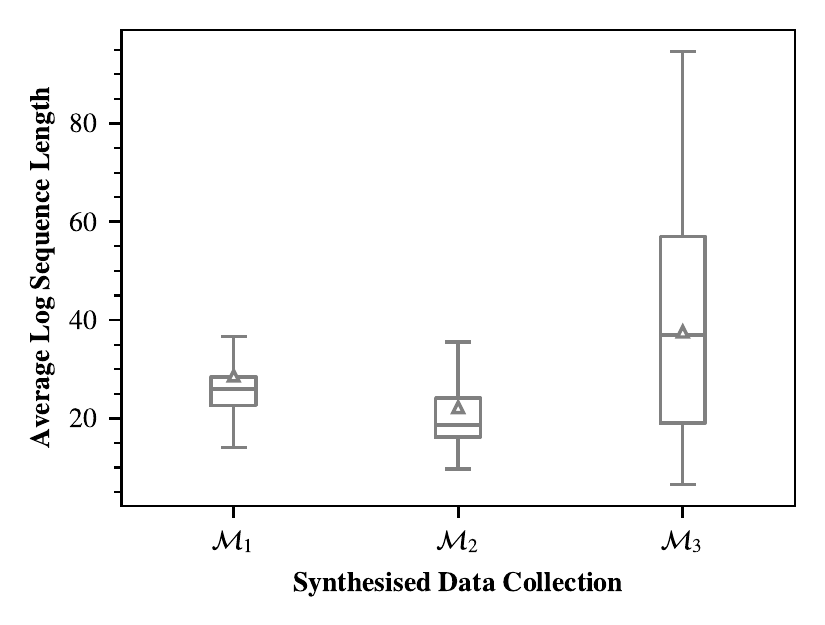}
    \caption{\label{fig:avgLSL}}
  \end{subfigure}%
  \hspace{0.3em}
  \begin{subfigure}{.45\linewidth}
    \centering
    \includegraphics[width = \linewidth]{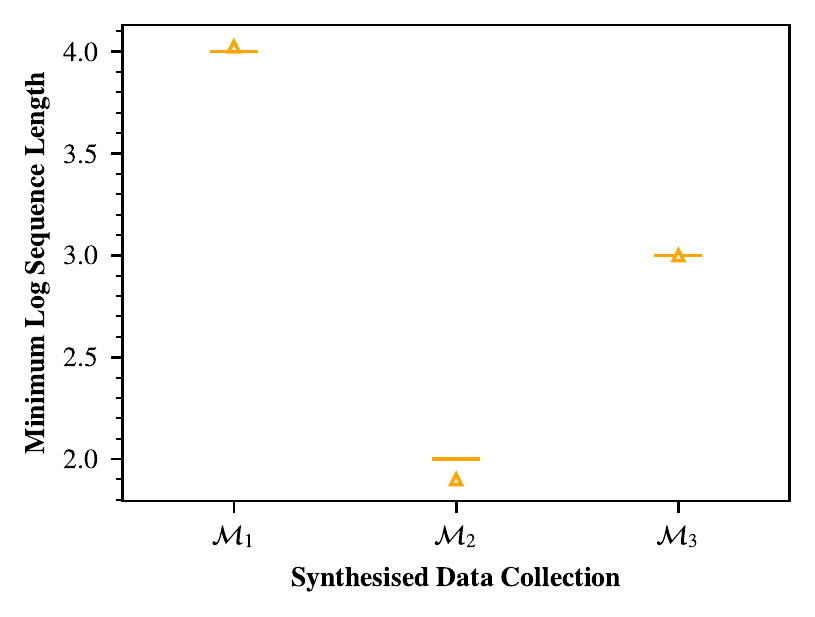}
    \caption{\label{fig:minLSL}}
  \end{subfigure}
    \hspace{0.2em}
    \begin{subfigure}{.45\linewidth}
    \centering
    \includegraphics[width = \linewidth]{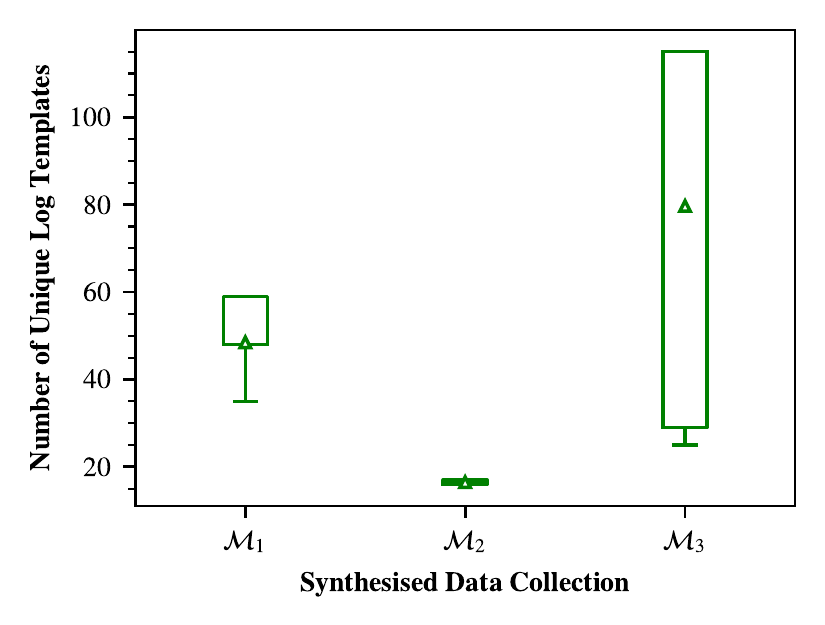}
    \caption{\label{fig:uniquelogtemplates}}
  \end{subfigure}%
  \hspace{0.3em}
\caption{Overview of Synthesised Datasets.\label{fig:datasetoverview}}
\end{figure}

As the correctness of the synthetic data generation was discussed in~\S~\ref{sec:syn-data-gen}, the synthesised datasets should follow the desired characteristics we specified  in~\S~\ref{sec:DatasetCharacteristics}. Here we present an overview of additional statistics regarding the dataset generation. 
\figurename~\ref{fig:datasetoverview} summarises three statistics in terms of average and minimum length of log sequences (Subfigure~\ref{fig:avgLSL} and Subfigure~\ref{fig:minLSL}) as well as the number of unique log templates in each dataset (Subfigure~\ref{fig:uniquelogtemplates}). Each box is based on 360 datasets generated from its corresponding behavioural model $\mathcal{M}_1$, $\mathcal{M}_2$, or $\mathcal{M}_3$. We note that log sequences inside synthesised datasets are directly generated by our approach introduced in~\S~\ref{sec:syn-data-gen}. Thus, no partitioning method is needed. However, given that each log sequence simulates a complete walk ( meaning from the initial state until the accepting state of a behavioural model), it more closely resembles the log sequences partitioned based on the log identifier. 
Based on Subfigure~\ref{fig:avgLSL}, synthesised datasets from $\mathcal{M}_3$ exhibit the largest interquartile ranges (IQRs), indicating a significant variation in average log sequence length. This arises from the higher complexity of $\mathcal{M}_3$ in terms of the number of states and templates (see~\S~\ref{sec:b-models}). Based on Subfigure~\ref{fig:minLSL}, the minimum length of log sequences remains relatively consistent across models. In Subfigure~\ref{fig:uniquelogtemplates}, M1 and M2 closely align with the number of unique templates reported for $\mathcal{M}_1$ and $\mathcal{M}_2$ in Table~\ref{table:models}. $\mathcal{M}_3$, however, shows a large IQR, ranging from the number of unique log templates in $\mathcal{M}_3$, 115, to as few as 35. 
Given that MLSL values (refer to Algorithm~\ref{alg:random-walk}) can be as low as 20, our algorithm is constrained to reach an accepting state within a specified number of transitions followed during a walk on the behavioural model, defined as a finite state automaton. In this way, there may be some transitions in the finite automaton that are not covered. Consequently, the log templates associated with the uncovered transitions are not present in the generated log sequence. 
Overall, the statistics of the synthesised datasets are consistent with our settings.

\section{Results}\label{sec:expr-results}
This section presents the results of RQ1 (DL encoders), RQ2 (log sequence embedding strategies), RQ3 (traditional ML), RQ4 (dataset characteristics), and RQ5 (real-world data), respectively. 

\subsection{RQ1: DL Encoders} \label{sec:res:rq1}

\begin{figure}
\centering
\includegraphics[width=0.6\textwidth]{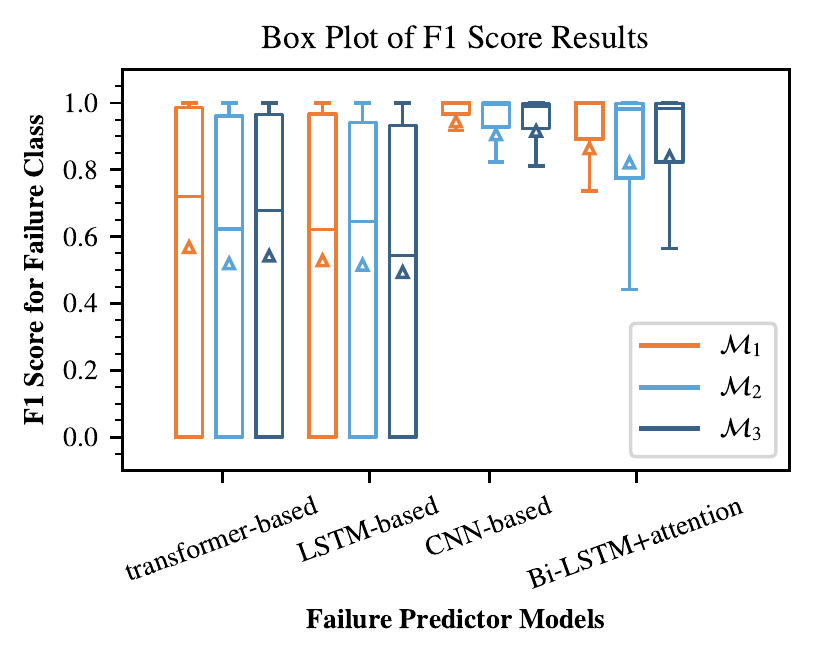}
\caption{Failure prediction accuracy for different DL encoders. The triangles additionally indicate mean values.}
\label{fig:rq1}
\end{figure}

\figurename~\ref{fig:rq1} shows boxplots of the failure prediction accuracy (F1 score) for different DL encoders (i.e., transformer-based, LSTM-based, CNN-based, and BiLSTM-based models) on the datasets generated by different behaviour models (i.e., $\mathcal{M}_1$, $\mathcal{M}_2$, and $\mathcal{M}_3$). 
Each box is generated based on $360\times 3$ data points since we have 360 combinations of dataset characteristics and three log sequence embedding strategies. 
In each box, a triangle indicates the mean value. 

Overall, the CNN-based model achieves the best performance in terms of F1 score for all behaviour models. It has the highest mean values with the smallest interquartile ranges (IQRs), meaning that the CNN-based model consistently works very well regardless of dataset characteristics and log sequence embedding strategies.  
The BiLSTM-based model also shows promising results. However, the CNN-based model's results are significantly higher for all the behavioural models (paired Wilcoxon test p-values $\ll$ 0.001). In contrast, the LSTM-based and transformer-based models show poor results (low F1 score on average with very large IQRs).
These patterns are independent from both the embedding strategy and the model. Further, the large IQR for LSTM-based and transformer-based models suggests that these models are very sensitive to the dataset characteristics.

The poor performance of the transformer-based encoder can be explained by the fact that the transformer blocks in the encoder are data-demanding (i.e., requiring much training data). When the dataset size is small (below \num{1000}), the data-demanding transformer blocks are not well-trained, leading to poor performance. This limitation is thoroughly discussed in the literature~\cite{trans-largedata}.

The LSTM-based encoder, on the other hand, has two simple layers of LSTM units. Recall that an LSTM model sequentially processes a given log sequence (i.e., a sequence of templates), template by template. Although LSTM attempts to address the long-term dependency problem of RNN by having a \emph{forget gate} (see \S~\ref{sec:background:rnn}), it is still a recurrent network that has difficulties to remember a long input sequence~\cite{Lipton2015ACR}. For this reason, since our log datasets contain long log sequences (up to a length of \num{1000}), the LSTM-based encoder did not work well. 

The BiLSTM-based encoder involves LSTM units and therefore has the weakness mentioned above. However, for BiLSTM, the input sequence flows in both directions in the network, utilizing information from both sides. Furthermore, it is enhanced by the attention mechanism that assigns more weight to parts of the input which are associated with the failure pattern~\cite{AttentionIA}.  Thus, the BiLSTM-based encoder can more easily learn the impact of different log templates on the classification results. However, the attention layer is more data-demanding than the convolution layers (see \S~\ref{sec:arch:DL:CNN}) in the CNN-based encoder, and this explains why the BiLSTM-based encoder does not outperform the CNN-based encoder.

The high performance of the BiLSTM-based and CNN-based encoders can be explained by the number of trainable parameters; for these two encoders, unlike the transformer-based and LSTM-based ones, the number of trainable parameters increases as the input sequences get longer. The larger number of parameters makes the encoders more robust to longer input log sequences.
Furthermore, CNN additionally processes spatial information (i.e., 
how templates relate to each other in the data) using multiple filters with different kernel sizes~\cite{CNNadvances}, which makes failure prediction more accurate even when the input size (sequence length) is large. 
These characteristics make the CNN-based encoder the best choice in our application context.

\begin{tcolorbox}
The answer to RQ1 is that the CNN-based encoder tends to significantly outperform the other encoders across the range of data characteristics and sequence embedding strategies.
\end{tcolorbox}

\subsection{RQ2: Log Sequence Embedding Strategies} \label{sec:res:rq2}

\begin{figure}
\centering
\includegraphics[width=0.6\textwidth]{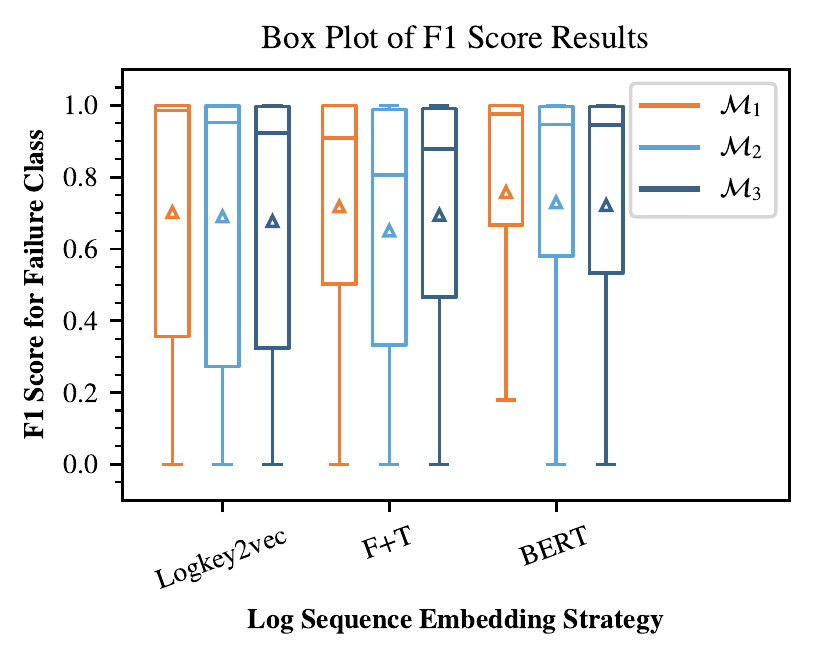}
\caption{Failure prediction accuracy for different log sequence embedding strategies. The triangles additionally indicate mean values.}
\label{fig:rq2-all}
\end{figure}

\figurename~\ref{fig:rq2-all} shows the boxplots of the failure prediction accuracy (F1 score) for the different log sequence embedding strategies considered in this study (i.e., BERT, F+T, and Logkey2vec) on the datasets generated by the three behaviour models ($\mathcal{M}_1$, $\mathcal{M}_2$, and $\mathcal{M}_3$). Each box is generated based on $360\times 4$ data points since we have \num{360} combinations of dataset characteristics and four DL encoders. Similar to \figurename~\ref{fig:rq1}, the triangle in each box indicates the mean value. We now inspect the plots shown inside with the aim of answering our research questions. The plots based on precision and recall are excluded since they draw similar conclusions.

\figurename~\ref{fig:rq2-all} shows that the BERT embedding strategy performs better than F+T and Logkey2vec for all behaviour models in terms of mean values and smaller IQRs. 
This means that, on average, for all DL encoders, the semantic-aware log sequence embedding using BERT fares better than both F+T, which employs FastText but is not as informative as BERT, and Logkey2vec, which solely relies on log template IDs and does not account for the semantic information of templates.

\begin{table}
    \centering
    \caption{Friedman test results (p-values). A level of significance $\alpha$ = 0.01 is used. In case of a significant difference, the best strategy is denoted as l (Logkey2vec), f (F+T), and b (BERT).\label{table:rq2-p-values}}
    \begin{tabular}{crrrr}
        \toprule
        \textbf{DL encoder}& \textbf{$\mathcal{M}_1$} & \textbf{$\mathcal{M}_2$} & \textbf{$\mathcal{M}_3$} & \textbf{All}\\
        \midrule
        CNN& $\ll 0.001$  (B)&  $\ll 0.001$ (L) &  $\ll 0.001$ (L) &  $\ll 0.001$ (L)\\
        BiLSTM & $\ll 0.001$ (B) & $\ll 0.001$ (B) & $0.001$ (B) & $\ll 0.001$ (B) \\
        transformer & $0.068$ & $\ll 0.001$ (F, B)& $\ll 0.001$ (F, B) & $\ll 0.001$ (F, B)\\
        LSTM & $\ll 0.001$ (B) & $\ll 0.001$ (L, B)& $\ll 0.001$ (L) & $\ll 0.001$ (B)\\
        \midrule
        \textbf{All} & $\ll 0.001$ (l)& $\ll 0.001$ (L, B)& $\ll 0.001$ (B) & $\ll 0.001$ (B)\\
        \bottomrule
    \end{tabular}
\end{table}

\begin{figure}
\centering
\includegraphics[width=0.6\textwidth]{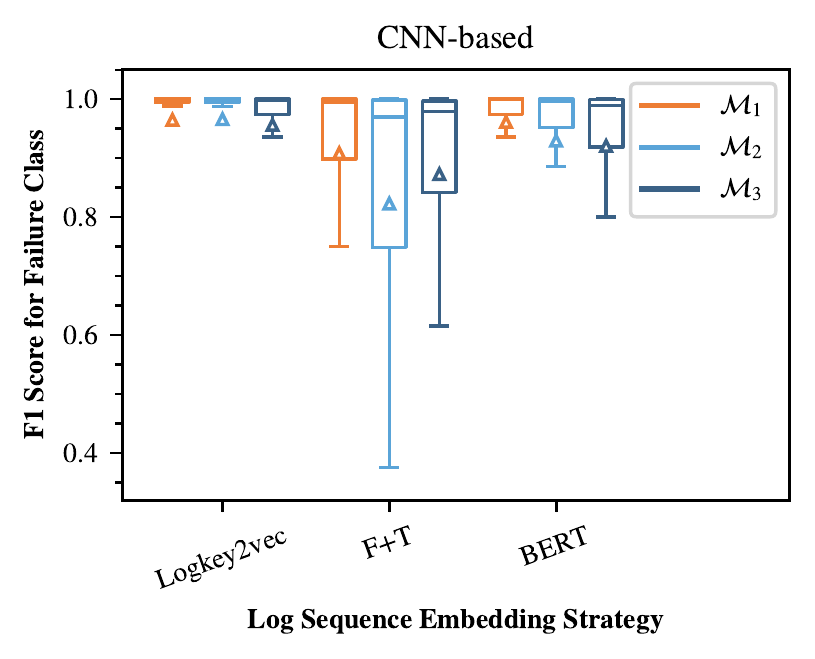}
\caption{Failure prediction accuracy of CNN-based model for different log sequence embedding strategies; triangles indicate mean values.}
\label{fig:rq2-cnn}
\end{figure}

To better understand the impact of log sequence embedding strategies on the performance of different DL encoders, we additionally performed Friedman test as a non-parametric test to compare the F1 score distributions of BERT, F+T, and Logkey2vec for each of the four DL encoders. Table~\ref{table:rq2-p-values} reports the statistical test results. For example, the low p-value in column \textit{$\mathcal{M}_2$} and row \textit{CNN} indicates that there are statistically significant differences between embedding strategies. In such cases, we employ a paired Wilcoxon test between each pair of embedding strategies and compare their medians to identify the top-performing strategy(ies). These are represented between brackets as L (Logkey2vec), F (F+T), and B (BERT) in the Table.

Interestingly, BERT is statistically better than or equal to F+T and Logkey2vec for all DL encoders except the CNN-based encoder (i.e., the best-performing DL encoder as investigated in \S~\ref{sec:res:rq1}) and the LSTM-based encoder for $\mathcal{M}_3$.
On the other hand, for the CNN-based encoder, the best overall embedding strategy is Logkey2vec, as clearly observable in  \figurename~\ref{fig:rq2-cnn}, depicting the F1 score distributions of Logkey2vec, F+T, and BERT for the CNN encoder.
In other words, combining the CNN-based encoder and the Logkey2vec embedding strategy is the best configuration of DL encoders and log sequence embedding strategies. 
Although, in contrast to BERT, Logkey2vec does not consider the semantic information of log templates, it accounts for the order of template IDs in each log sequence. Furthermore, Logkey2vec is trained together with the DL encoder, while BERT is pre-trained independently from the DL encoder. We suspect that such characteristics of Logkey2vec play a positive role when combined with the CNN-based encoder.
F+T presents the largest IQR and lowest mean and median. This observation is consistent with the overall strategy comparison depicted in Fig~\ref{fig:rq2-all}, and similar rationales apply.

We note that BERT is still an attractive strategy for log sequence embedding when any other encoder than CNN is used. Although BERT is considerably larger than Logkey2vec in terms of parameters, using BERT does not require significantly more time and resources than Logkey2vec and F+T since BERT minimises repeated calculations by mapping each log template to its corresponding BERT embedding vector.

\begin{tcolorbox}
The answer to RQ2 is that the performance of the log sequence embedding strategies varies depending on the DL encoders used. Although BERT outperforms F+T and Logkey2vec overall across all encoders, Logkey2vec outperforms BERT when the CNN-based encoder is used. 
\end{tcolorbox}

\subsection{RQ3: Traditional ML}\label{sec:res:rq3}
\begin{figure}[t]
\centering
\includegraphics[width=0.6\textwidth]{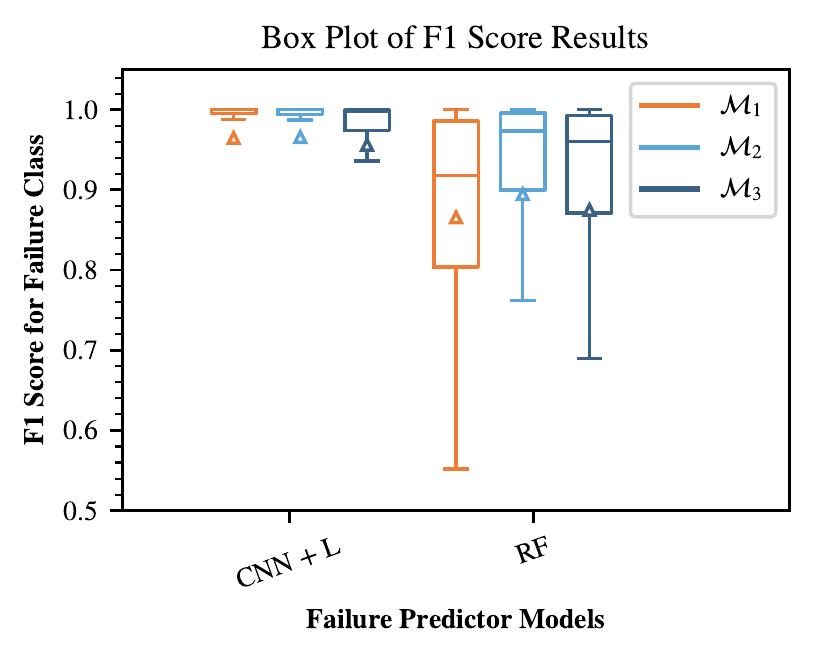}
\caption{Failure prediction accuracy of the best DL-based configuration (CNN with Logkey2vec) next to a traditional ML-based configuration (RF); triangles depict mean values.}
\label{fig:rq3-cnnvsrf}
\end{figure}

\figurename~\ref{fig:rq3-cnnvsrf} shows the boxplots of the failure prediction accuracy (F1 score) for the best configuration of the DL encoder and the log sequence embedding strategy, i.e., the CNN-based encoder and Logkey2vec, next to one of the best performing~\cite{Fernandez-Delgado2014, prefix, rfvsxgboost}, traditional ML-based failure predictor (RF), on the datasets generated by the three behaviour models ($\mathcal{M}_1$, $\mathcal{M}_2$, and $\mathcal{M}_3$). Each box is generated based on $360$ data points since we have \num{360} combinations of dataset characteristics. Similar to the previous boxplots, the triangle in each box indicates the mean value. We shall now examine the provided plots aiming to address our research question.

In \figurename~\ref{fig:rq3-cnnvsrf}, the CNN-based encoder with Logkey2vec clearly achieves significantly higher accuracy and robustness compared to RF, in terms of average accuracy and IQR, respectively, regardless of the behaviour models used to generate the log datasets. 
RF relies on aggregating decisions from multiple trees which can limit its ability to capture intricate, non-linear patterns of failures. In contrast, CNNs, as described in \S~\ref{sec:background:cnn}, use convolutional layers to automatically extract hierarchical features from embedded representations, combined with pooling layers that reduce spatial dimensions, allowing CNNs to handle more complex patterns.
Additionally, as explained in \S~\ref{sec:methodology}, the input of RF is an embedding vector rather than an embedding matrix using TF-IDF, which is a template-ID-based strategy. The best DL configuration also uses a template ID-based strategy, Logkey2vec. However, unlike TF-IDF, Logkey2vec embeddings keep updating during failure predictor training; this enables logkey2vec to learn the embeddings with respect to labels of the log sequences, see~\S~\ref{sec:background:logkey2vec}.

\begin{tcolorbox}
The answer to RQ3 is that, using the best configuration of the DL-based failure predictor, i.e., the CNN-based encoder and Logkey2vec, results in significantly higher accuracy and robustness (low IQR) compared to Random Forest, which is considered one of the top traditional ML classifiers.
\end{tcolorbox}

\subsection{RQ4: Dataset Characteristics} \label{sec:res:rq4}

Recall that there are 12 possible configurations for the DL-based architecture (i.e., four DL encoders and three embedding strategies), each of which may exhibit varying performances across different data set characteristics. Although CNN+L (CNN-based encoder with Logkey2vec) is the best configuration overall based on RQ1 and RQ2 results, there may be datasets where other configurations fare better. 
Therefore, it could potentially be informative to investigate each of the configurations in terms of their accuracy for different dataset characteristics.
However, many configurations clearly provide low accuracy for most of the datasets and do not significantly outperform the other cases. 
So we first determined the best configurations worth investigating across the 1080 datasets. 
Specifically, for each configuration, we counted the number of datasets for which that configuration is among the best.
We defined a threshold $r$ set to 0.01 to include all configurations with a difference in accuracy value lower than the threshold $r$. This way, we could account for all high-performing configurations.
It turned out that only the following three configurations kept appearing among the best configurations for almost all datasets\footnote{There were only 72 out of 1080 datasets where the configurations other than the top three configurations were among the best. However, not only these were very rare but their accuracy was too low to be useful.}: CNN+L (CNN encoder with Logkey2vec), CNN+B (CNN-based encoder with BERT), and BiLSTM+B (BiLSTM-based encoder with BERT). 
Note that the top three configurations remained the same for different threshold values ($r$ = 0, 0.05, 0.1). 
Table~\ref{tab:rq4-outlier} provides more details about the three best configurations; column ``\#Best ($r=0.01$)'' provides the number of instances where the confuguration is among the best, and additional columns ``Avg'', ``Med'', ``Min'', and ``Max'' show the average, median, minimum, and maximum F1 scores for the configurations, respectively.
Based on the above observations, we focus our analysis of dataset characteristics on the three best configurations, while providing the same plots for the rest of the configurations as supplementary material in our replication package (see \S~\ref{sec:data-availablility}).

\begin{table*} 
    \caption{Overview of the Three Best Configurations for DL-based Failure Prediction}\label{tab:rq4-outlier} 
    \centering

    \begin{threeparttable}[t]
    \begin{tabular}{crrrrrr}
        \toprule
        \textbf{Rank} 
        &  \textbf{\#Best ($r=0.01$)} & \textbf{Config} & \textbf{Avg} & \textbf{Med} & \textbf{Min} & \textbf{Max}\\
        \midrule
        1 & 866 &  CNN+L & 0.962 & 1.0 & 0.0 & 1.0\\
        2 &  667 & CNN+B & 0.936 & 0.997 & 0.0 & 1.0\\
        3 &  627 & BiLSTM+B & 0.879 & 0.995 & 0.0 & 1.0\\
        \bottomrule
    \end{tabular}
\end{threeparttable}
\end{table*}

\begin{figure*}
\begin{subfigure}{.45\linewidth}
    \centering
    \includegraphics[width = \linewidth]{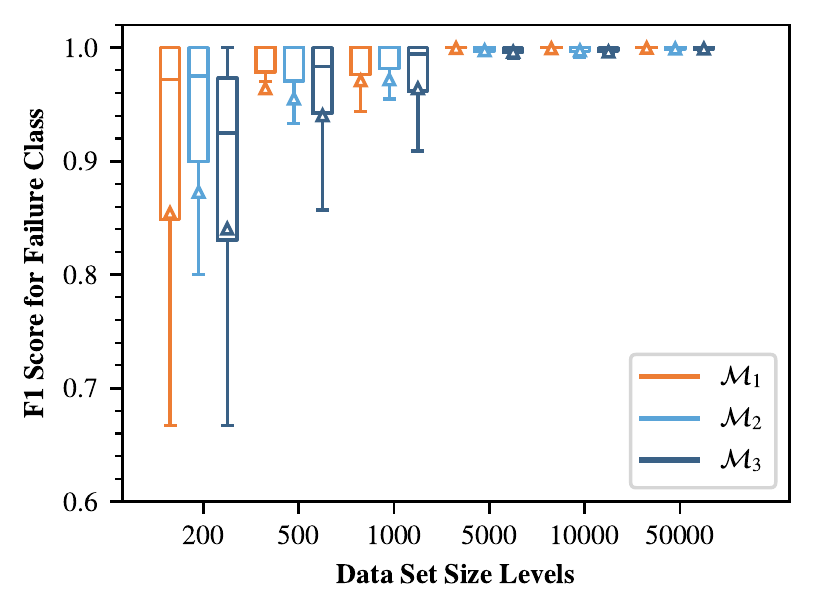}
    \caption{\label{fig:factorsboxplot-a}}
  \end{subfigure}%
  \hspace{0.3em}
  \begin{subfigure}{.45\linewidth}
    \centering
    \includegraphics[width = \linewidth]{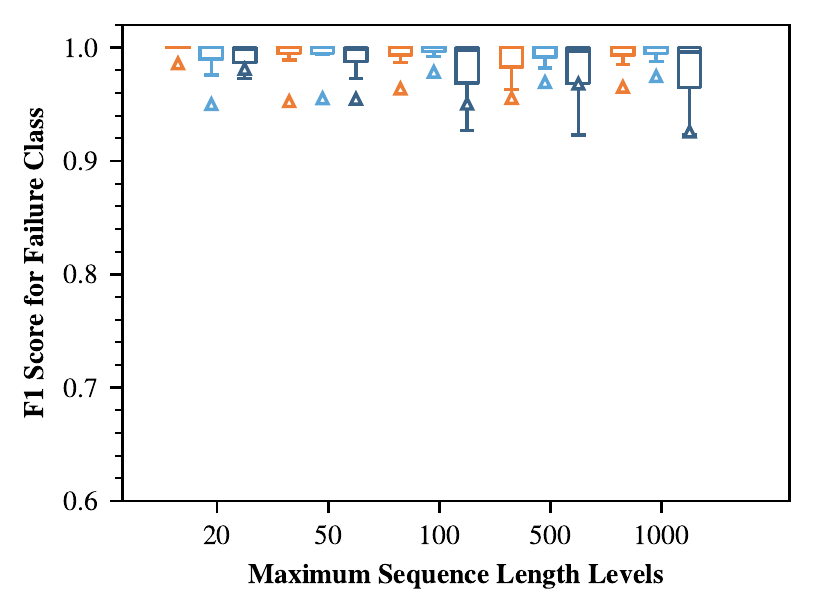}
    \caption{\label{fig:factorsboxplot-b}}
  \end{subfigure}
    \hspace{0.2em}
    \begin{subfigure}{.45\linewidth}
    \centering
    \includegraphics[width = \linewidth]{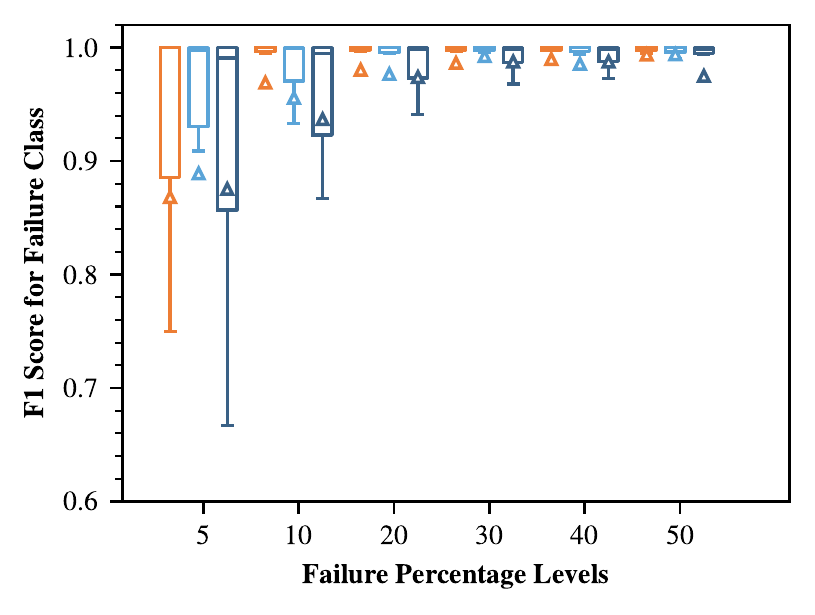}
    \caption{\label{fig:factorsboxplot-c}}
  \end{subfigure}%
  \hspace{0.3em}
   \begin{subfigure}{.45\linewidth}
    \centering
    \includegraphics[width = \linewidth]{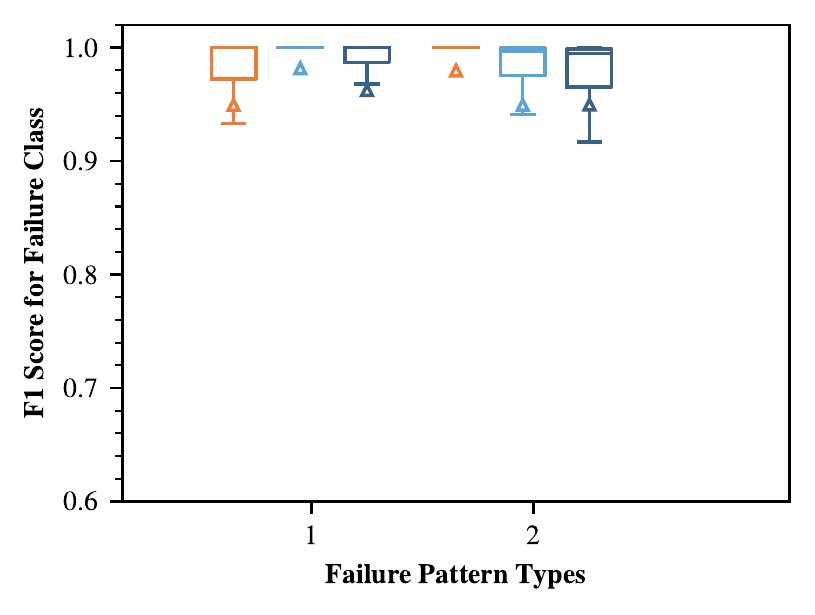}
    \caption{\label{fig:factorsboxplot-d}}
  \end{subfigure}%
  \hspace{0.3em}

 \caption{Failure prediction accuracy of the CNN-based encoder with Logkey2vec for different dataset characteristics}
 \label{fig:factorsboxplot}
\end{figure*}

\begin{figure*}
\centering
     \includegraphics[scale=0.7]{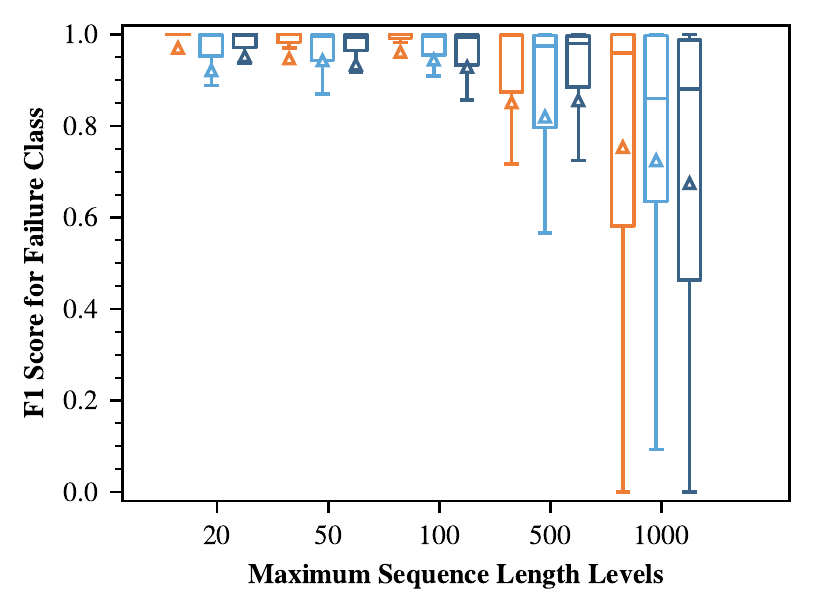}
 \caption{Failure prediction accuracy of the BiLSTM-based encoder with BERT as a function of maximum sequence length}
 \label{fig:rq4-bi-b}
\end{figure*}

\figurename~\ref{fig:factorsboxplot} shows the distributions of F1 scores according to different dataset characteristic values for CNN+L, the best configuration overall. To save space, we have excluded the plots for the second-best and third-best configurations from the paper as they were very similar to CNN+L, except for the maximum sequence length for BiLSTM+B, which will be discussed separately later. However, all the remaining plots can be found in our replication package, as previously mentioned. 
We discuss next how the failure prediction accuracy of CNN+L varies with each of the dataset characteristics.

In \figurename~\ref{fig:factorsboxplot-a}, we can see the impact of dataset size on the failure prediction accuracy; it is clear that accuracy decreases with smaller datasets, regardless of the behaviour models used to generate the log datasets. For example, when dataset size is \num{200}, accuracy decreases below \num{0.7} in the worst case, whereas it always stays very close to \num{1.0} when dataset size is above or equal to \num{5000}. Since larger datasets imply more training data, this result is intuitive but it clarifies data requirements for failure prediction. 

\figurename~\ref{fig:factorsboxplot-b} depicts the impact of maximum LSL values (\textit{MLSL}) on the failure prediction accuracy. Compared to the impact of data set size, we can see that its impact is relatively small. This implies that  CNN+L works fairly well for long log sequence lengths of up to \num{1000}. We suspect that the impact of log sequence length could be significant for much longer log sequences. However, log sequences longer than \num{1000} are not common in  publicly available, real-world log datasets~\cite{10.1145/3510003.3510155} as explained in Section~\ref{sec:DatasetCharacteristics}. Nevertheless, the investigation of much longer log sequences would be informative.

The relationship between failure percentage and failure prediction accuracy (F1 score) is depicted in \figurename~\ref{fig:factorsboxplot-c}. It is clear that, overall, the F1 score increases as the failure percentage increases. This is intuitive since a larger failure percentage means more instances of failure patterns in the training data, making it easier to learn such patterns. 
An interesting observation is that the average failure prediction accuracy is above \num{0.9} even when the failure percentage is \num{10}\%. This implies that CNN+L can cope well with unbalanced data.

\figurename~\ref{fig:factorsboxplot-d} shows the failure prediction accuracy for different failure pattern types. There is no consistent trend across models $\mathcal{M}_1$, $\mathcal{M}_2$, and $\mathcal{M}_3$; \textit{Type-F} (the corresponding language is finite) is easier to detect than \textit{Type-I} (the corresponding language is infinite) in $\mathcal{M}_2$ and $\mathcal{M}_3$, whereas the opposite happens in $\mathcal{M}_1$. It is unclear why, in $\mathcal{M}_1$, detecting less complex failure patterns (\textit{Type-F}) is more difficult than detecting more complex patterns (\textit{Type-I}). We may not have defined failure pattern types in a way that is conducive to explaining variations in accuracy and different hypotheses will have to be tested in future work with respect to which pattern characteristics matter.

As mentioned earlier, BiLSTM+B shows a distinct result only for longer log sequences, as depicted in \figurename~\ref{fig:rq4-bi-b}.
Unlike CNN+L shown in \figurename~\ref{fig:factorsboxplot}(b), larger IQR and lower average values are clearly visible for longer log sequences in \figurename~\ref{fig:rq4-bi-b}. This indicates that significantly increasing the maximum length of log sequences decreases the failure prediction accuracy of BiLSTM+B.

\begin{figure}
\centering
\centering
\begin{tikzpicture}[
    node distance = 5mm and 7mm,
    module/.style={%
        draw, rounded corners,
        minimum width=#1,
        minimum height=5mm,
        font=\linespread{1}\selectfont
        },
    module/.default=2cm,
    >=LaTeX,
 disc/.style = {shape=cylinder, draw, shape aspect=0.27,
                shape border rotate=90,
                text width=20mm, align=center, font=\linespread{1}\selectfont},
  mdl/.style = {shape=ellipse, aspect=3, draw},
  alg/.style = {draw, align=center, font=\linespread{1}\selectfont},
  alg2/.style = {draw, align=center, 
  minimum width=#1,
        minimum height=5mm,
        font=\linespread{1}\selectfont},
        alg2/.default=2cm
                    ]
    \node [module = 2cm] (n1) {\small \textit{Dataset size $<=$ 3000}};
    \node [module= 2cm, align = center, fill = green!10!white, below right = 5 mm and -8 mm of n1 ] (n2) 
    {\small \textit{CNN+L} \\ 540 samples \\ 0.998 avg F1 \\ 1.0 med F1\\ 0.962 min F1 \\ 1.0 max F1};
    \node [module= 2cm, below left = 5 mm and -8 mm of n1 ] (n3) 
    {\small \textit{Failure Percentage $<=$ 15}};
    \node [module= 2cm,align = center, fill = orange!10!white, below left = 5 mm and -2 mm of n3 ] (n4) 
    {\small \textit{BiLSTM+B} \\ 180 samples \\ 0.717 avg F1 \\ 0.857 med F1\\ 0.0 min F1 \\ 1.0 max F1};
    \node [module= 2cm, below right = 7 mm and -19 mm of n3 ] (n5) 
    {\small \textit{Dataset Size $<=$ 350}};
    \node [module= 2cm,align = center, fill = green!10!white, below right = 5 mm and -8 mm of n5 ] (n6) 
    {\small \textit{CNN+L} \\120 samples \\ 0.975 avg F1\\ 1.0 med F1\\0.0 min F1 \\ 1.0 max F1};
    \node [module= 2cm,align = center, fill = blue!10!white, below left = 5 mm and -8 mm of n5 ] (n7) 
    {\small \textit{CNN+B} \\ 240 samples \\ 0.877 avg F1\\ 0.909 med F1\\0.286 min F1 \\ 1.0 max F1};

    \draw [->] (n1) --node[above right = -1mm and 3 mm] {No}(n2); 
    \draw [->] (n1) --node[above left = -0.5mm and 2 mm] {Yes}(n3); 
    \draw [->] (n3) --(n4);
    \draw [->] (n3) --(n5);
    \draw [->] (n5) --(n6); 
    \draw [->] (n5) --(n7);
\end{tikzpicture}
\caption{Decision Tree identifying the best configurations based on dataset characteristics\label{fig:rq4:decisiontree}}
\end{figure}
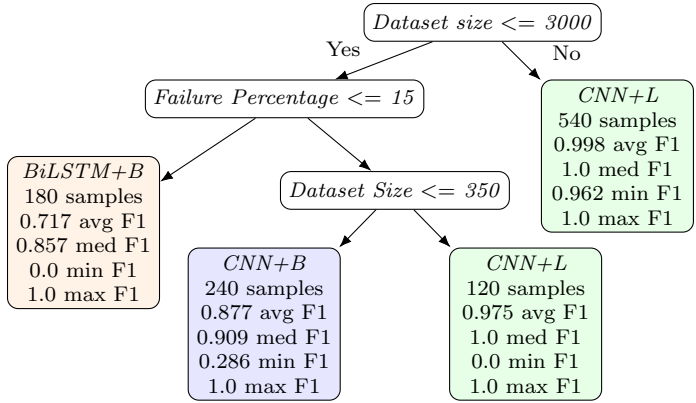

To further investigate the data set characteristics that work well with the three best configurations, we built a classification tree predicting the best configuration for given dataset characteristics. 
To do this, we first labelled the 1080 datasets with the three best configurations; specifically, each dataset was labelled with the top performer among the three configurations. 
We then split the 1080 datasets into subsets of 720 (66.7\%) and 360 (33.3\%) datasets for training and testing the classification tree, respectively. 
Since the training data was imbalanced due to the superior performance of CNN+L for most datasets, we applied higher weights to minority classes using Inverse Proportional Weighting~\cite{He2009LearningFI} to address the class imbalance issue.
We also performed Minimal Cost-Complexity Pruning (MCCP)~\cite{BreiFrieStonOlsh84} to avoid over-fitting.
\figurename~\ref{fig:rq4:decisiontree} shows the resulting classification tree, where each non-leaf node captures a decision condition and each leaf node the (predicted) best configuration for the conditions corresponding to the path from the root to the leaf. Each leaf node also includes the number of samples in the leaf, as well as the average (``avg''), median (``med''), minimum (``min''), and maximum (``max'') F1 score for the predicted configuration. 
For example, the right-most leaf node indicates that CNN+L is the best configuration when the dataset size is larger than 3000. A total of 540 of the 1080 datasets satisfy this condition, and we can expect a failure prediction accuracy of 0.998 when using CNN+L.

The classification tree shows that dataset size and, to a lesser extent, failure percentage play a pivotal role in determining the best configuration for the DL-based failure predictor.
Specifically, CNN+L is recommended for dataset sizes larger than 3000. However, for smaller dataset sizes, if the failure percentage is lower than or equal to 15\%, BiLSTM+B is the recommended configuration. 
In other words, for dataset sizes lower than or equal to 3000 and failure percentages lower than or equal to 15\%, BiLSTM+B performs better than CNN+L and CNN+B. 
We suspect this result is due to BiLSTM+B's higher capability in the presence of highly imbalanced datasets. 
In contrast, if the failure percentage is above 15\%, CNN+B is recommended when the dataset size is lower than or equal to 350, while CNN+L is recommended when the dataset size is higher than 350. 
In other words, for dataset sizes lower than or equal to 3000 and failure percentages higher than 15\%, CNN+B performs the best. 
This can be attributed to the challenges posed by a small dataset for training  logkey2vec from scratch, leading to better semantic-enabled embeddings from BERT.

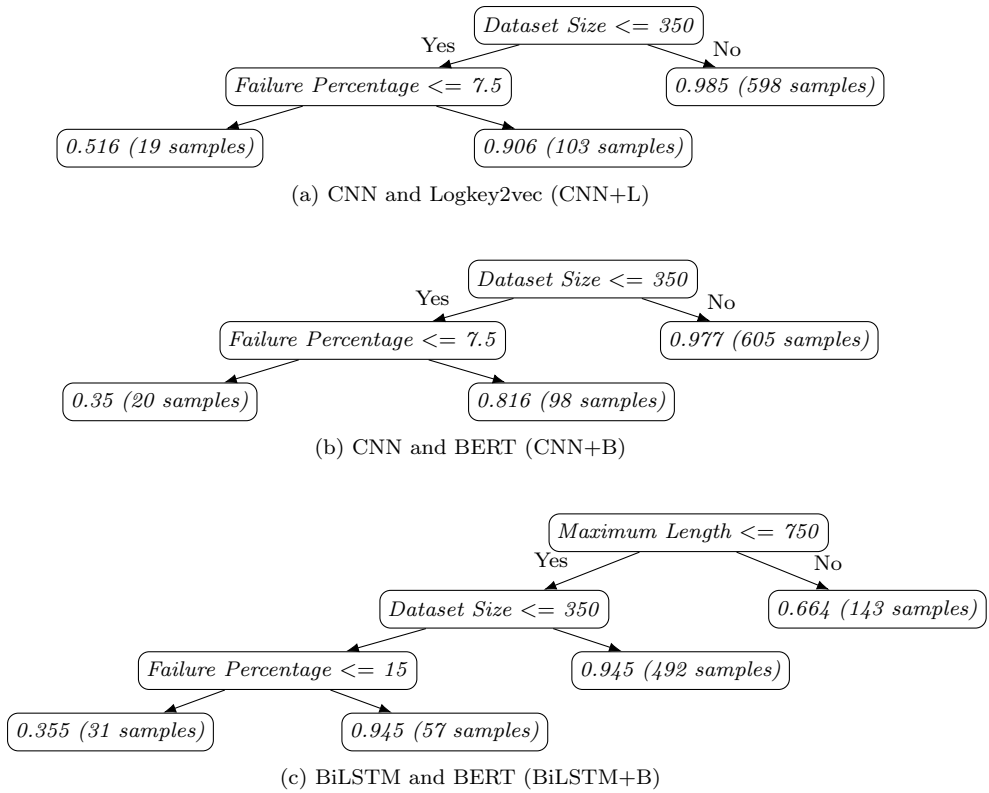
\begin{figure}
\begin{subfigure}{\linewidth}
    \centering
\begin{tikzpicture}[
    node distance = 5mm and 7mm,
    module/.style={%
        draw, rounded corners,
        minimum width=#1,
        minimum height=5mm,
        font=\linespread{1}\selectfont
        },
    module/.default=2cm,
    >=LaTeX,
 disc/.style = {shape=cylinder, draw, shape aspect=0.27,
                shape border rotate=90,
                text width=20mm, align=center, font=\linespread{1}\selectfont},
  mdl/.style = {shape=ellipse, aspect=3, draw},
  alg/.style = {draw, align=center, font=\linespread{1}\selectfont},
  alg2/.style = {draw, align=center, 
  minimum width=#1,
        minimum height=5mm,
        font=\linespread{1}\selectfont},
        alg2/.default=2cm
                    ]
    \node [module = 2cm] (n1) {\small \textit{Dataset Size $<=$ 350}};
    \node [module= 2cm, below right = 3 mm and -5 mm of n1 ] (n2) 
    {\small \textit{0.985 (598 samples)}};
    \node [module= 2cm, below left = 3 mm and -5 mm of n1 ] (n3) 
    {\small \textit{Failure Percentage $<=$ 7.5}};
    \node [module= 2cm, below right = 3 mm and -5 mm of n3 ] (n4) 
    {\small \textit{0.906 (103 samples)}};
    \node [module= 2cm, below left = 3 mm and -5 mm of n3 ] (n5) 
    {\small \textit{0.516 (19 samples)}};

    \draw [->] (n1) --node[above right = -1mm and 3 mm] {No}(n2); 
    \draw [->] (n1) --node[above left = -0.5mm and 2 mm] {Yes}(n3); 
    \draw [->] (n3) --(n4); 
    \draw [->] (n3) --(n5); 
\end{tikzpicture}
    \caption{CNN and Logkey2vec (CNN+L)\label{fig:rq3:regressiontree}}
  \end{subfigure}%
  \hspace{2em}
  \newline
  \newline

   \begin{subfigure}{\linewidth}
    \centering
\begin{tikzpicture}[
    node distance = 5mm and 7mm,
    module/.style={%
        draw, rounded corners,
        minimum width=#1,
        minimum height=5mm,
        font=\linespread{1}\selectfont
        },
    module/.default=2cm,
    >=LaTeX,
 disc/.style = {shape=cylinder, draw, shape aspect=0.27,
                shape border rotate=90,
                text width=20mm, align=center, font=\linespread{1}\selectfont},
  mdl/.style = {shape=ellipse, aspect=3, draw},
  alg/.style = {draw, align=center, font=\linespread{1}\selectfont},
  alg2/.style = {draw, align=center, 
  minimum width=#1,
        minimum height=5mm,
        font=\linespread{1}\selectfont},
        alg2/.default=2cm
                    ]
    \node [module = 2cm] (n1) {\small \textit{Dataset Size $<=$ 350}};
    \node [module= 2cm, below right = 3 mm and -5 mm of n1 ] (n2) 
    {\small \textit{0.977 (605 samples)}};
    \node [module= 2cm, below left = 3 mm and -5 mm of n1 ] (n3) 
    {\small \textit{Failure Percentage $<=$ 7.5}};
    \node [module= 2cm, below right = 3 mm and -5 mm of n3 ] (n4) 
    {\small \textit{0.816 (98 samples)}};
    \node [module= 2cm, below left = 3 mm and -5 mm of n3 ] (n5) 
    {\small \textit{0.35 (20 samples)}};

    \draw [->] (n1) --node[above right = -1mm and 3 mm] {No}(n2); 
    \draw [->] (n1) --node[above left = -0.5mm and 2 mm] {Yes}(n3); 
    \draw [->] (n3) --(n4); 
    \draw [->] (n3) --(n5);
\end{tikzpicture}
    \caption{CNN and BERT (CNN+B)\label{fig:rq4:reg-cn-b}}
    \end{subfigure}
    \hspace{0.2em}
    \newline
    
  \begin{subfigure}{\linewidth}
    \centering
\begin{tikzpicture}[
    node distance = 5mm and 7mm,
    module/.style={%
        draw, rounded corners,
        minimum width=#1,
        minimum height=5mm,
        font=\linespread{1}\selectfont
        },
    module/.default=2cm,
    >=LaTeX,
 disc/.style = {shape=cylinder, draw, shape aspect=0.27,
                shape border rotate=90,
                text width=20mm, align=center, font=\linespread{1}\selectfont},
  mdl/.style = {shape=ellipse, aspect=3, draw},
  alg/.style = {draw, align=center, font=\linespread{1}\selectfont},
  alg2/.style = {draw, align=center, 
  minimum width=#1,
        minimum height=5mm,
        font=\linespread{1}\selectfont},
        alg2/.default=2cm
                    ]
    \node [module = 2cm] (n1) {\small \textit{Maximum Length $<=$ 750}};
    \node [module= 2cm, below right = 5 mm and -8 mm of n1 ] (n2) 
    {\small \textit{0.664 (143 samples)}};
    \node [module= 2cm, below left = 5 mm and -8 mm of n1 ] (n3) 
    {\small \textit{Dataset Size $<=$ 350}};
    \node [module= 2cm, below right = 3 mm and -5 mm of n3 ] (n4) 
    {\small \textit{0.945 (492 samples)}};
    \node [module= 2cm, below left = 3 mm and -5 mm of n3 ] (n5) 
    {\small \textit{Failure Percentage $<=$ 15}};
    \node [module= 2cm, below right = 3 mm and -10 mm of n5 ] (n6) 
    {\small \textit{0.945 (57 samples)}};
    \node [module= 2cm, below left = 3 mm and -10 mm of n5 ] (n7) 
    {\small \textit{0.355 (31 samples)}};

    \draw [->] (n1) --node[above right = -1mm and 3 mm] {No}(n2); 
    \draw [->] (n1) --node[above left = -0.5mm and 2 mm] {Yes}(n3); 
    \draw [->] (n3) --(n4); 
    \draw [->] (n3) --(n5);
    \draw [->] (n5) --(n6); 
    \draw [->] (n5) --(n7);
\end{tikzpicture}
    \caption{BiLSTM and BERT (BiLSTM+B)\label{fig:rq4:reg-bi-b}}
  \end{subfigure}

    \caption{Regression Tree for the best configurations based on dataset characteristics in  F1 scores}\label{fig:rq4:reg}
\end{figure}

We additionally built regression trees for each of the three best configurations to further investigate how their failure prediction accuracy varies according to dataset characteristics. 
We applied the same approach used for pruning the classification tree above.

\figurename~\ref{fig:rq4:reg} depicts the regression trees for CNN+L (\figurename~\ref{fig:rq3:regressiontree}), CNN+B (\figurename~\ref{fig:rq4:reg-cn-b}), and BiSLTM+B (\figurename~\ref{fig:rq4:reg-bi-b}). 
For example, in \figurename~\ref{fig:rq3:regressiontree}, the left-most leaf node indicates that the average failure prediction accuracy is predicted to be \num{0.516} if the dataset size is less than or equal to \num{350} \textit{and} the failure percentage is less than or equal to \num{7.5}. Otherwise, the failure average prediction accuracy is predicted to be \num{0.9}. 

From the regression trees, it is clear that dataset size and failure percentage are once again the two main factors that explain variations in failure prediction accuracy. 
Both CNN+L and CNN+B show similar results: the accuracy decreases significantly when the dataset size is less than or equal to 350 and the failure percentage is less than or equal to 7.5. 
BiLSTM+B also exhibits low accuracy in similar conditions (i.e., when both the dataset size and the failure percentage are small), but it additionally shows a low accuracy when the maximum log sequence length is higher than 750. 

More practical implications and guidelines derived from the classification and regression trees will be further discussed in Section~\ref{sec:discus:findings}.

\begin{tcolorbox}
The answer to RQ4 is that dataset size, followed by failure percentage, plays an important role in the accuracy of DL-based failure predictors while LSL is important only for some configurations. In contrast, failure pattern type does not have a clear relationship with failure prediction accuracy. 

Interestingly, failure predictors are very accurate ($\textit{F1-score} > 0.95$) and robust ($IQR < 0.01$) when dataset size is above 350 or failure percentage is above 7.5\%. 
\end{tcolorbox}

\subsection{RQ5: Real-world Data}\label{sec:res:rq5}

Table~\ref{table:rq5} shows the accuracy results of synthesized datasets alongside those of the real-world dataset, OpenStack\_FP, for the best the same DL-based configuration, CNN+L. The ``Dataset'' column lists the datasets chosen for comparison. Synthesised$_{\mathcal{M}_1}$, Synthesised$_{\mathcal{M}_2}$, and Synthesised$_{\mathcal{M}_3}$ are the datasets generated from the $\mathcal{M}_1$, $\mathcal{M}_2$, and $\mathcal{M}_3$ behavioural models, with similar characteristics to OpenStack\_FP in terms of dataset size, maximum log sequence length, and percentage of failure, denoted by ``DS'', ``MLSL'', and ``PF'', respectively. ``CNN+L'' stands for the most effective configuration based on RQ1-3 results. 
For each behavioural model, there are two dataset instances matching these three characteristics but having different failure pattern types (\textit{Type-F} and \textit{Type-I}). The values of precision, recall, and F1 score (denoted by ``P'', ``R'', and ``F1'', respectively) are shown under the ``CNN+L'' column, for a more detailed comparison. Since we do not have information regarding the failure pattern types of OpenStack\_PF, the table presents an average of the two synthesised datasets in each row dedicated to Synthesised data. Furthermore, the fourth row shows the average accuracy results from all synthesised behavioural models.

\begin{table}[htbp]
    \centering
    \caption{Comparison of Results from a Real-world Dataset (OpenStack\_FP) with Synthesised Datasets with similar characteristics}\label{table:rq5}
    \begin{tabular}{crrrrrr}
        \toprule
        \multirow{2}{*}{\textbf{Dataset}} & \multirow{2}{*}{\textbf{DS}} & \multirow{2}{*}{\textbf{MLSL}} & \multirow{2}{*}{\textbf{PF}} & \multicolumn{3}{c}{\textbf{CNN + L}}\\
        & & & & P & R & F1\\ 
        \midrule
        Synthesised$_{\mathcal{M}_1}$ & 1000 & 500 & 20 & $0.987$ & $1.000$ & $0.993$ \\
        Synthesised$_{\mathcal{M}_2}$ & 1000 & 500 & 20 & $0.932$ & $0.965$ & $0.9480$\\
        Synthesised$_{\mathcal{M}_3}$ & 1000 & 500 & 20 & $0.947$ & $0.988$ & $0.967$\\
        \cline{5-7}\addlinespace
        average &  &  &  & $0.955$ & $0.984$ & $0.969$ \\
         \midrule
        OpenStack\_FP & 876 & 468 & 21.46 & $0.974$ & $0.974$ & $0.974$\\
        \bottomrule
    \end{tabular}
\end{table}

According to Table~\ref{table:rq5}, the average F1 score for synthesised datasets shows a difference below $0.01$ with OpenStack\_FP ($0.969$ vs $0.974$). The precision values obtained on the synthesised datasets are slightly lower than those obtained on OpenStack\_FP, while the recall values are slightly higher. The average difference amounts to $0.019$ for precision and $0.010$ for recall.  
To rigorously assess the significance of this difference, we performed the Wilcoxon test between the accuracy results obtained on synthesised data and those obtained on OpenStack\_PF; for each test, the accuracy results from the synthesized datasets were paired with the results from OpenStack\_PF. All p-values for precision, recall, and F1 score are far above 0.05, indicating that the differences between real-world and synthesized datasets are statistically insignificant. 

\begin{tcolorbox}
The answer to RQ5 is that there is no significant difference between the accuracy results obtained on comparable synthesised datasets and a real-world one (OpenStack\_FP) when using the best configuration for failure prediction (CNN-based encoder with Logkey2vec).
\end{tcolorbox}

\subsection{Data Availability Statement}\label{sec:data-availablility}

The replication package, including the implementation, generated datasets with behavioural models, and results, is publicly available~\cite{replication-package}.

\section{Discussion}\label{sec:discussion}
\subsection{Findings and Implications}\label{sec:discus:findings}

Our study leverages the main DL types (LSTM, CNN, and transformer), along with all categories of LSE strategies (Logkey2vec, BERT, and hybrid strategy of FastText and TF-IDF). In contrast to other studies mentioned in \tableautorefname~\ref{tab:relatedworks}, the full configuration of DL encoders and LSE strategies are evaluated. Moreover, instead of using a limited number of datasets, using synthesized data enables us to control dataset characteristics to identify necessary conditions for achieving high-accuracy models. 
Nonetheless, we also considered a real-world dataset for failure prediction (OpenStack\_PF) and applied it to the best failure predictor configuration. This allows us to compare the failure prediction accuracy results obtained on the synthesized datasets with those obtained on the OpenStack FP dataset.

Several major findings are reported in \S~\ref{sec:expr-results}. 
First, the CNN-based DL encoder fares the best among different DL encoders, including the ones based on LSTMs, transformers, and BiLSTMs.
Second, the CNN-based DL encoder works best with the Logkey2vec embedding strategy, although BERT fares better than Logkey2vec and the hybrid of FastText and TF-IDF overall for all DL encoders.
Third, compared to the leading traditional ML approaches, such as Random Forest, the best DL-based failure predictor configuration yields significantly higher accuracy and robustness.
Fourth, although the CNN-based DL encoder and the Logkey2vec embedding strategy are not the most recent techniques in their respective fields, interestingly, their configuration (CNN+L) works best overall for failure prediction. For CNN+L, both the size and the failure percentage of input log datasets significantly drive the failure prediction accuracy, whereas the log sequence length and the failure pattern type do not. Similar trends have been observed for the second-best configuration (CNN+B). However, for the third-best configuration (BiLSTM+B), besides the above relations, MLSL increasing the maximum length of log sequences significantly
decreases the failure prediction accuracy.

Fifth, based on the analysis of~\S~\ref{sec:res:rq4}, we are able to provide comprehensive guidelines. In general, 
for datasets larger than 3000, CNN+L is the recommended configuration. Conversely, when dataset sizes are 3000 or less and the dataset's failure percentage is at most 15\%, the preferred choice is BiLSTM+B.
Regarding the expected accuracy, the accuracy of both CNN+L and CNN+B significantly reduces if the dataset size is 350 or below, and the failure percentage is up to 7.5\%. While BiLSTM+B accuracy is directly affected by the maximum log sequence length, accuracy further decreases when it exceeds 750. If the maximum log sequence is at most 750, BiLSTM+B significantly decreases when the dataset size is 350 or below, similar to CNN+L and CNN+B, and the failure percentage is up to 15\%.

The conditions driving failure prediction accuracy suggest practical guidelines. For example, for a log dataset size below \num{350} and a failure percentage below \num{7.5}\%, failure prediction using CNN+L will be inaccurate and cannot be trusted. In that case, one can increase either the log dataset size or the failure percentage to build a better failure predictor. Although the failure percentage is inherent to the system under analysis and might not be easy to control in practice, collecting more log sequences during the operation of the system to increase the dataset size is usually feasible.

Last but not least, using the best configuration, the accuracy results obtained on synthesised and real-world datasets do not present a significant difference, hence further suggesting our data synthesis approach is valid.

Below we discuss the practical implications of our findings for the main stakeholders: AIOps engineers and software engineering researchers.

\paragraph{AIOps Engineers.} Proactive maintenance is an important part of AIOps engineering~\cite{aiops}. Failure prediction is therefore a crucial part of alleviating the impact of failures. In this study, the analysis of the best configurations of the failure prediction model described in~\S~\ref{sec:res:rq2} can guide engineers in choosing the most appropriate options when designing an architecture for their data. Our guidelines, based on the decision and regression trees presented in~\S~\ref{sec:res:rq3}, narrow the scope of possible design choices by decreasing the number of candidate configurations based on the characteristics of the dataset. 
Furthermore, we remark that the implementation of our modular architecture is available (see~\S~\ref{sec:data-availablility}), enabling AIOps engineers to reuse our artifacts seamlessly.
\paragraph{Software Engineering Researchers.} In this paper, we use a modular architecture to effectively study different DL architectures on failure prediction data. Since existing approaches apply DL models with selective settings such as LSE strategies~\cite{Das2018, aarohi}, we propose a novel approach to study configurations of LSE strategies and DL architectures that have not been studied together before (see Table~\ref{tab:relatedworks}). We speculate this approach can further inspire the adaptation of DL-based modular architectures in other studies in the field of AI for software engineering. In addition, we use a controllable synthetic data generation algorithm to generate labeled datasets with varying characteristics. Such datasets are crucial to obtaining comprehensive and generalisable results when only a limited number of datasets are available for assessing a new method. We believe the algorithm presented in~\S~\ref{sec:syn-data-gen} can be adopted to generate synthetic datasets tailored to specific requirements.

\subsection{Threats to Validity}\label{sec:threats}
There are a number of potential threats to the validity of our experimental results.

\paragraph{Hyper-parameter tuning of models.}
The hyper-parameters of failure predictors, such as optimizers, loss functions, and learning rates, can affect the results. To mitigate this, we followed recommendations from the literature. For the batch size and the number of epochs, as mentioned in \S~\ref{sec:training-testing}, we chose values for different combinations of dataset characteristics based on preliminary evaluation results. Better results could be obtained with different choices. 

\paragraph{Synthetic data generation process.} 
Due to the lack of a method to generate the datasets satisfying different dataset characteristics mentioned in \S~\ref{sec:data-generator-reqs}, we proposed a new approach, with precise algorithms, that can generate datasets in a controlled, unbiased manner as discussed in \S~\ref{sec:syn-data-gen}. To mitigate any risks related to synthetic generation, we provided proof of the correctness of the algorithms and explained why it is unbiased during the generation process in \S~\ref{sec:syn-correctness}. To further support the validity of the generation process, in \S~\ref{sec:res:rq5}, we compared the results on actual datasets reported in the literature with those of the synthesised datasets for corresponding key parameters (e.g., dataset sizes and failure percentage). Results show to be remarkably consistent, thus backing up the validity of our experiments. 

\paragraph{Timeliness of failure predictions.} 
Depending on the context, the timeliness of failure prediction may impact the applicability of our DL models. Because the focus of our experiments is on prediction accuracy, we have not investigated how early our DL models can accurately predict failures; we simply predict failures after processing all log messages (up to the moment before the failure message occurs) within a log sequence. 
Investigating timeliness would require entirely different experiments; for example, this can be done by varying the distance between the last log message inputted to DL models and the occurrence of failures, either in terms of the number of log messages or time difference. 
However, due to the objective and design of our study, we use the entire log sequence before the failure for prediction, meaning the distance between the last log message in the observation window inputted to DL models and the failure log message is zero \emph{by design}.
We remark that for the real-world OpenStack\_FP dataset, which contains timestamps, the average time distance between the last message before failure and the failure message is \SI{1.87}{\s}. However, interpreting whether such a lapse is sufficient in practice requires to know the practical context in which the prediction models are deployed. 
We acknowledge the limitations of our datasets and the need to study the timeliness of failure prediction for DL models systematically in the future. 

\paragraph{Behavioural models and failure patterns.} 
The behavioural models and failure patterns used for the generation of synthetic datasets may have a significant impact on the experimental results. 
We want to remark that this is the first attempt to characterise failure patterns for investigating failure prediction performance. 
To mitigate this issue, we carefully chose them based on pre-defined criteria described in \S~\ref{sec:expr-setting} and provided a remark on its generalizability in \S~\ref{sec:Generalisability}.
Nevertheless, more case studies, especially considering finer-grained failure patterns, are required to increase the generalizability of our findings and implications and, for that purpose, we provide in our replication package all the artifacts required.

\paragraph{Possible bugs in the implementation.}
The implementation of the DL encoders, the log sequence embedding strategies, the dataset generation algorithms, and the scripts used in our experiments might include unexpected bugs. To mitigate this risk, we used the replication packages of existing studies~\cite{experimentreport, Neurallog} as much as possible. Also, we carefully performed code reviews.

\section{Conclusion}\label{sec:conclusion}

In this paper, we presented a comprehensive and systematic evaluation of alternative failure prediction strategies relying on DL encoders and log sequence embedding strategies. We presented a generic, modular architecture for failure prediction which can be configured with specific DL encoders and embedding strategies, resulting in different failure predictors. We considered Logkey2vec, BERT, and a hybrid of FastText and TF-IDF, representing three categories of log sequence embedding strategies. We also covered the main DL categories resulting in four DL encoders (LSTM-, BiLSTM-, CNN-, and transformer-based). Our selection was inspired by the previously used DL models in the literature. 

We evaluated the failure prediction models on diverse synthetic datasets using three behavioural models inferred from available system logs. Four dataset characteristics were controlled when generating datasets: dataset size, failure percentage, Log Sequence Length (LSL), and failure pattern type. Using these characteristics, 360 datasets were generated for each of three behavioural models.

Evaluation results show that the accuracy of the CNN-based encoder is significantly higher than the other encoders regardless of dataset characteristics and embedding strategies. Among the three embedding strategies, pretrained BERT outperformed Logkey2vec and the hybrid strategy overall, although Logkey2vec fared better for the CNN-based encoder. Compared to the best traditional ML-based failure predictor (Random Forest), the best configuration demonstrates significantly superior accuracy and robustness.
The analysis of dataset characteristics confirms that increasing the dataset size and failure percentage increases failure prediction accuracy. In comparison, LSL is a significant factor only for specific configurations, while the other factors (i.e., failure pattern type) did not show a clear relationship with accuracy. Furthermore, the accuracy of the best configuration (i.e., CNN-based with Logkey2vec) consistently yielded high accuracy when the dataset size was above \num{350} \textit{or} the failure percentage was above \num{7.5}\%, which makes it widely usable in practice. 
Finally, the accuracy results obtained from the synthesized and real datasets are consistent.

As part of future work, we plan to further evaluate the best-performing configurations of the failure prediction architecture on additional real-world log data to further investigate the effect of other factors, such as log parsing techniques or data noise, on model accuracy. 
As future research direction, we plan to assess the impact of more dataset characteristics on log-based failure prediction. This notably includes different sources of data noises such as varying degrees of mislabelled logs, log
parsing errors, and evolving logs. The degree and type of data noise are, however, dependent on the system of study and such noise may not be significant on all datasets.
Finally, following the discussion on the timeliness of failure prediction and limitations of our datasets in~\S~\ref{sec:threats}, when using real-world data, we also plan to include additional evaluation metrics, such as lead time~\cite{Salfner2010} and the number of log messages before the occurrence of a failure, to assess the accuracy of models at predicting failures early on. 

\textbf{Acknowledgements. }This work was supported by the Canada Research Chair and Discovery Grant programs of the Natural Sciences and Engineering Research Council of Canada (NSERC), by a University of Luxembourg’s joint research program grant, the Science Foundation Ireland under Grant 13/RC/2094-2, and by European Union's Horizon 2020 Research and Innovation Programme under grant agreement No. 957254 (COSMOS).
The experiments conducted in this work were enabled in part by Digital Alliance of Canada  (alliancecan.ca).

\section*{Declarations}


\textbf{Funding and/or Conflicts of interests/Competing interests} The authors declare that they have no conflict of interest.

\bibliographystyle{spbasic}
\bibliography{refer}

\vspace{10pt}
\noindent 
\begin{minipage}[t]{0.25\textwidth}
  \vspace{0pt} 
  \includegraphics[width=\linewidth]{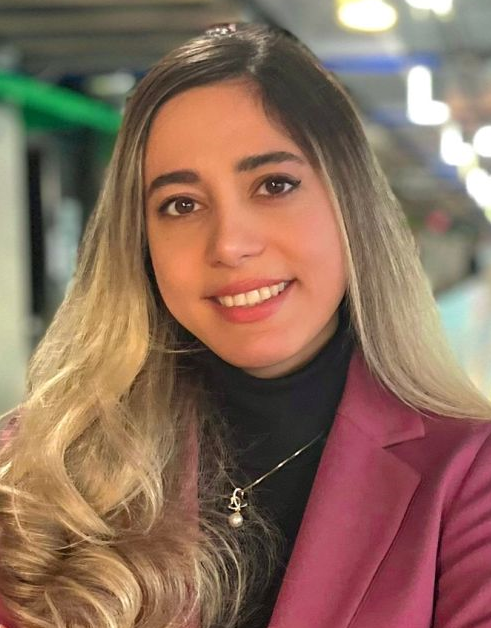} 
\end{minipage}%
\hfill 
\begin{minipage}[t]{0.7\textwidth}
  \vspace{0pt} 
  \small \noindent \textbf{Fatemeh Hadadi} is currently working toward a Ph.D. degree under the supervision of Prof. Lionel Briand with the School of EECS, University of Ottawa, and a member of Nanda Lab. In 2021, she was awarded the prize of first class in computer science from the supporter foundation of the University of Tehran and was granted a direct Ph.D. from B.C. by the University of Ottawa. Her research interests include log analysis (e.g., failure prediction and anomaly detection), natural language processing, and applied machine learning. 
\end{minipage}

\vspace{10pt}

\noindent 
\begin{minipage}[t]{0.25\textwidth}
  \vspace{0pt} 
  \includegraphics[width=\linewidth]{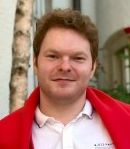} 
\end{minipage}%
\hfill 
\begin{minipage}[t]{0.7\textwidth}
  \vspace{0pt} 
  \small \noindent \textbf{Joshua H. Dawes} completed his PhD at CERN (Geneva, Switzerland)
and was awarded the final degree by the University of Manchester
(Manchester, UK) in 2021. During his time at CERN, Josh developed
and applied formal verification techniques for software that served
operations of the Large Hadron Collider. After CERN he spent time at
SnT in Luxembourg where, while he was helping with this paper, his
formal verification software was applied by companies in the Avionics
and Acoustics sectors. Having concluded his stint in academia, Josh is
now a Senior Software Engineering Manager at the ISIS Neutron and
Muon Source in the UK.
 
\end{minipage}

\vspace{10pt}
\noindent 
\begin{minipage}[t]{0.25\textwidth}
  \vspace{0pt} 
  \includegraphics[width=\linewidth]{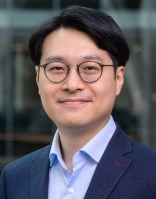} 
\end{minipage}%
\hfill 
\begin{minipage}[t]{0.7\textwidth}
  \vspace{0pt} 
  \small \noindent \textbf{Donghwan Shin} did his BS, MSc, and PhD at Korea Advanced Institute of Science and Technology (KAIST), South Korea. This was followed by four years as a research associate/scientist at the SVV (Software Verification and Validation) group, the Interdisciplinary Centre
for ICT Security, Reliability, and Trust (SnT) of the University of
Luxembourg. He is currently a lecturer (assistant professor in the
American system) at the Department of Computer Science, University of Sheffield, UK. His research and teaching interests lie in testing
for ML-enabled cyber-physical systems (e.g., ML-enabled automated
driving systems), log analysis (e.g., model inference and anomaly
detection), and mutation testing. More details can be found at \url{https://dshin.info}. 
\end{minipage}

\vspace{10pt}

\noindent 
\begin{minipage}[t]{0.25\textwidth}
  \vspace{0pt} 
  \includegraphics[width=\linewidth]{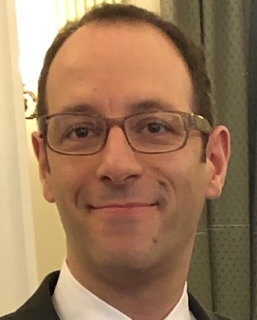} 
\end{minipage}%
\hfill 
\begin{minipage}[t]{0.7\textwidth}
  \vspace{0pt} 
  \small \noindent \textbf{Domenico Bianculli} is associate professor/chief scientist 2 at the Interdisciplinary
  Centre for Security, Reliability and Trust (SnT), University of
  Luxembourg.  He holds a PhD degree from Università della Svizzera
  italiana (Lugano, Switzerland), a MSc in Computing Systems
  Engineering and a BSc in Computer Engineering, both from Politecnico
  di Milano (Milan, Italy).  Domenico's research focuses on the
  specification and verification of evolvable software systems. His
  research interests include: run-time verification, temporal logics and specification languages, log analysis,
  program analysis, and regulatory compliance. 
\end{minipage}
\vspace{10pt}

\noindent 
\begin{minipage}[t]{0.25\textwidth}
  \vspace{0pt} 
  \includegraphics[width=\linewidth]{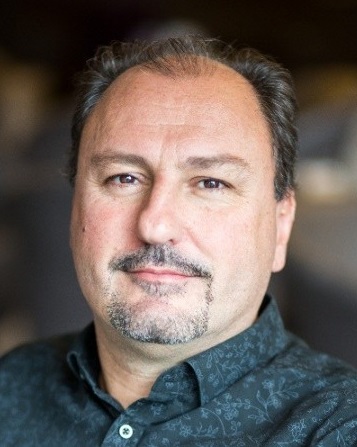} 
\end{minipage}%
\hfill 
\begin{minipage}[t]{0.7\textwidth}
  \vspace{0pt} 
  \small \noindent \textbf{ Lionel C. Briand} is professor of software engineering and has shared appointments between (1) The University of Ottawa, Canada, and (2) The Lero SFI Centre---the national Irish centre for software research---hosted by the University of Limerick, Ireland. In collaboration with colleagues, for over 30 years, he has run many collaborative research projects with companies in the automotive, satellite, aerospace, energy, financial, and legal domains. Lionel has held various engineering, academic, and leading positions in seven countries.  He currently holds a Canada Research Chair (Tier 1) on "Intelligent Software Dependability and Compliance" and is the director of Lero, the national Irish re for software research. Lionel was elevated to the grades of IEEE Fellow and ACM Fellow for his work on software testing and verification. Further, he was granted the IEEE Computer Society Harlan Mills award, the ACM SIGSOFT outstanding research award, and the IEEE Reliability Society engineer-of-the-year award. He also received an ERC Advanced grant in 2016 on modelling and testing cyber-physical systems, the most prestigious individual research award in the European Union and was elected a fellow of the Academy of Science, Royal Society of Canada in 2023. More details can be found at: \url{http://www.lbriand.info}. 
\end{minipage}

\end{document}